\documentclass[twocolumn]{aastex631}

\usepackage{graphicx}
\usepackage{dcolumn}

\usepackage{mathtools}
\usepackage{bm}
\usepackage{xcolor}

\begin{document}

\title{Hierarchical Inference of Binary Neutron Star Mass Distribution and Equation of State with Gravitational Waves}

\author[0000-0002-6977-670X]{Jacob Golomb}
\email{jgolomb@caltech.edu}
\affiliation{LIGO Laboratory, California Institute of Technology, Pasadena, CA 91125, USA}
\author[0000-0003-2053-5582]{Colm Talbot}
\affiliation{LIGO Laboratory, California Institute of Technology, Pasadena, CA 91125, USA}
\affiliation{LIGO Laboratory, Massachusetts Institute of Technology, Cambridge, Massachusetts 02139, USA}
\affiliation{Kavli Institute for Astrophysics and Space Research, Massachusetts Institute of Technology, Cambridge, Massachusetts 02139, USA}

\date{\today}

\begin{abstract}
Gravitational wave observations of binary neutron star mergers provide valuable information about neutron star structure and the equation of state of dense nuclear matter. Numerous methods have been proposed to analyze the population of observed neutron stars and previous work has demonstrated the necessity of jointly fitting the astrophysical distribution and the equation of state in order to accurately constrain the equation of state. In this work, we introduce a new framework to simultaneously infer the distribution of binary neutron star masses and the nuclear equation of state using Gaussian mixture model density estimates which mitigates some of the limitations previously-used methods suffer from. Using our method, we reproduce previous projections for the expected precision of our joint mass distribution and equation of state inference with tens of observations. We also show that mismodeling the equation of state can bias our inference of the neutron star mass distribution. While we focus on neutron star masses and matter effects, our method is widely applicable to population inference problems.
\end{abstract}

\section{Introduction}

Over the past six years, the LIGO-Virgo gravitational wave detectors~\citep{LIGO, Virgo}  have made $> 50$ observations of black hole and neutron star binary mergers~\citep{gwtc2-catalog}, providing a new way to study some of the most energetic events in the universe.
Growing catalogs of gravitational wave observations allow us to study the populations from which compact binary systems originate, offering further insights into the physical nature of these systems \citep{gwtc1, gwtc2, Zackay2019b, Zackay2019, Venumadhav2020, OGC2020}. 
 
Population inference from gravitational wave observations is performed by comparing catalogs of observed events to models of the astrophysical distribution.
These astrophysical models include strongly physically motivated models (e.g.,~\cite{Zevin2021, Wong2021}), phenomenological models inspired by theoretical predictions and prior observations (e.g.,~\cite{farrow2019, Wysocki2020}), or data-driven models (e.g.,~\cite{Tiwari2021}).

Such population studies are an example of hierarchical Bayesian inference, combining a set of observed events, marginalizing over the single-event parameters for each event, and extracting global properties that govern the single-event parameters to probe the underlying distribution of events, putting observed properties of single events into a wider astrophysical context (see, e.g. \citet{Thrane:2018qnx, Mandel2019, vitale2020} for recent reviews.)
Using gravitational wave observations in a hierarchical framework provides a powerful method of constraining universal properties of merging binary neutron star (BNS) systems, such as the BNS mass distribution \citep{Farr2011, farrow2019} and equation of state (EOS) of dense nuclear matter \citep{Agathos2015, Tsui:2005zf, Wysocki2020, Lackey2015, Hernandez2019, ghosh2021,Landry2019, ChatziioannouFarr2020}.

While terrestrial experiments have constrained the EOS of cold nuclear matter for densities approaching the nuclear saturation density, a complete picture of the microphysics of nuclear matter above these densities has yet to be confidently determined.
With central densities reaching several times nuclear saturation density, neutron stars \textemdash observed via kilonovae spectra and light curves, X-ray pulsar measurements, and gravitational waves \textemdash provide a probe of nuclear physics at super-saturation densities \citep{Bogdanov2019, Miller2019, Miller2021, Silva2021,Metzger2019, Coughlin2018}.

The first detection of a BNS merger via gravitational waves \citep{gw170817Discovery, gw170817} provided constraints on the EOS of dense nuclear matter, favoring more ``soft'' or compressible EOSs over stiffer EOSs \citep{eos170817}.
The second observed binary neutron star merger \citep{gw190425} did not provide significant constraints on the EOS; however, the relatively high mass of this system suggests a tension with the galactic population of binary neutron star systems \citep{gw190425, Galaudage2021}. 

While only two confident detections of BNS mergers have been made by LIGO-Virgo, further detections in the near future will provide constraints on the EOS of high-density nuclear matter through the combination of observed events \citep{margalit2019, Chatziioannou2020}.
Unlike gravitational wave signals from binary black hole mergers, signals from BNS mergers contain information about the neutron star EOS. This information is primarily encoded by the tidal deformabilities of the two bodies during late-stage inspiral, with the magnitude of this effect determined by the underlying EOS \citep{Hinderer2010, Bildsten1992, Zhao2018}.

Specifically, the EOS directly governs the pressure-density relationship inside the star, a necessary ingredient for solving the Tolman-Oppenheimer-Volkoff equations for the mass-radius relationship of neutron stars \citep{Zhao2018, Lindbloom1992}. For a given EOS, the mass determines the magnitude of a neutron star's quadrupole moment induced from the external field during merger, imprinting a signature in the detected gravitational wave signal \citep{Hinderer2010, Chatziioannou2018, Chatziioannou2020, gw170817, eos170817, Thorne1998}. This imprint is commonly expressed in terms of dimensionless tidal deformability ($\Lambda$), which is defined as \citep{Chatziioannou2020, eos170817, Flanagan2008}
\begin{equation}
    \Lambda = \frac{\lambda}{m^5} = \frac{2}{3} k_2 {\left(\frac{R}{m}\right)}^5,
\end{equation}
where $k_2$ is the quadrupole Love number, $R$ is the radius of the neutron star, and $m$ is its mass (we express this formula in units where $c = G = 1$). The EOS determines both $k_2$ and $R$ for a given neutron star mass, resulting in a unique $\Lambda - m$ relationship for different (hadronic) EOSs \citep{Zhao2018, Chatziioannou2020, Hinderer2010, wade2014}. Under the assumption of a common EOS among neutron stars, we can infer this $\Lambda - m$ relationship by combining observations, constraining the underlying EOS.

Previous work has shown that hierarchical inference can be used to constrain the neutron star EOS by considering parameterized models of neutron star populations in conjunction with EOS relations \citep{Lackey2015, Landry2019, Wysocki2020}. In \cite{Wysocki2020}, the authors emphasize the importance of simultaneously inferring the mass distribution and EOS, due to bias that results from independent analyses.
In this work, we introduce and implement a new method of performing a simultaneous hierarchical analysis to infer mass distribution and EOS.
Specifically, we use a Gaussian mixture model (GMM) as an estimate of single-event posterior probability densities.
Using this method, we demonstrate that mismodeling the EOS can lead to a biased inference of the neutron star mass distribution.

The paper is organized as follows.
In Section \ref{Methods}, we detail the process of our density estimation procedure and how it can be implemented in general hierarchical inference problems.
We then outline our choice of parameterized BNS mass population and EOS models in Section \ref{Models}.
We follow this in Section \ref{Data} with the details of the simulated data we use for the proof-of-concept and a description of how we apply this method to simultaneous EOS and mass distribution inference.
We review the results of our simulated data study in Section \ref{Results} and conclude with takeaways and motivations for future work in Section \ref{Discussion}.

\section{Methods}\label{Methods}

We begin by reviewing Bayesian inference in the context of gravitational wave data analysis (see, e.g., \cite{Thrane:2018qnx, vitale2020} for recent reviews).
In Bayesian inference, one constructs the posterior distribution $p(\Theta|d)$ for a model with parameters $\Theta$ given some data $d$. Bayes' theorem is typically written as
\begin{equation}\label{bayes}
    p(\Theta|d) = \frac{\mathcal{L}(d|\Theta)\pi(\Theta)}{Z(d)}
\end{equation}
where $\mathcal{L}(d|\Theta)$ is the likelihood of the data given the model parameters, $\pi(\Theta)$ is the prior probability distribution, characterizing our prior beliefs on the distribution of $\Theta$, and $Z(d)$ is the Bayesian evidence, or the marginal likelihood for the model \footnote{While Equation \ref{bayes} is technically conditioned on a model, we suppress the explicit dependence in our notation.}.
In gravitational wave analysis, $\mathcal{L}(d|\Theta)$ is typically taken to be a Gaussian likelihood distribution, whose mean is given by a (frequency domain) gravitational waveform characterized by $\Theta$ and variance given by the detector noise (e.g.,~\cite{veitch2015}).
The full set of $\Theta$ typically contains parameters intrinsic to the merger event (such as masses and spins) as well as the extrinsic parameters, such as position in the sky and luminosity distance. 

Because the set of parameters $\Theta$ is typically $> 10$ dimensional, to recover the posterior distribution, Equation \ref{bayes} is commonly sampled iteratively using a Markov Chain Monte Carlo (MCMC) sampler or nested sampler (e.g.,~\cite{veitch2015, bilby}).
Once the sampler converges, we are left with a set of samples drawn from the posterior distribution.
This posterior distribution represents our full knowledge about the physical parameters of the source of the gravitational wave event with our prior distribution.

Now, we combine multiple events hierarchically and sample the \textit{hyper}posterior, in order to learn about the \textit{hyper}parameters (or population parameters) $\Omega$ that describe the global distribution of a subset of single-event parameters (e.g., the distribution of neutron star masses). We denote these single-event parameters of interest as $\theta$, a vector of length $D$, which is a subset of $\Theta$. 

We do the combination by replacing the fixed model above with the set of hyperparameters describing the population model \citep{Thrane:2018qnx, vitale2020}.

\begin{equation}\label{h_bayes}
    p(\Omega|\mathbf{d}) = \frac{\pi(\Omega)\mathcal{L}(\mathbf{d}|\Omega)}{Z(\mathbf{d})}.
\end{equation}
In the above notation, $\pi(\Omega)$ is the hyper-prior, and $Z$ is the Bayesian evidence for all the observed data $\mathbf{d}$ marginalized over the hyper-prior.
Assuming the observed events are $N$ independent draws from the population, we express the (hyper-) likelihood as
\begin{equation}\label{hyperlikelihood}
    \mathcal{L}(\mathbf{d}|\Omega) = \prod^{N}_{i=1} \int \mathcal{L}(d_i|\theta_i) p(\theta_i|\Omega)d\theta_i ,
\end{equation}
where for each observed event we marginalize over the single-event parameters $\theta$ conditioned on a population model $p(\theta|\Omega)$.
The likelihood $\mathcal{L}(d_i|\theta)$ is implicitly already marginalized over all members of $\Theta$ not included in $\theta$.

In order to account for selection effects, we augment Equation \ref{hyperlikelihood} by including a selection term:
\begin{equation}\label{pdet}
    P_{\text{det}}(\Omega) = \frac{1}{N_{\textrm{inj}}} \sum^{N_\textrm{found}}_{i=1} \frac{p(\theta_i|\Omega)}{p(\theta_i|\Omega_0)}.
\end{equation}

We compute this term by injecting into simulated detector noise $N_{\textrm{inj}}$ simulated events from a fiducial source population $\Omega_0$, and determining how many of those events pass our SNR detection threshold (see Section~\ref{Data}).
In this equation, the term $\theta_i$ consists of the parameters of the $i$th ``found'' injection. Our new total likelihood is:
\begin{equation}\label{likelihood}
        \mathcal{L}(\mathbf{d}|\Omega) = P_{\text{det}}(\Omega)^{-N} \prod^{N}_{i=1} \int \mathcal{L}(d_i|\theta_i) p(\theta_i|\Omega)d\theta_i.
\end{equation}

A key challenge in population inference is efficiently evaluating Equation \ref{hyperlikelihood} for a large catalog of events as Equation \ref{h_bayes} is typically also constructed using stochastic sampling, requiring as many as several million evaluations. 

Current techniques to compute this integral via Monte Carlo integration involve reweighting posterior samples (assuming the fiducial prior) by the corresponding population likelihood for each sample, for example in the analysis performed in~\cite{gwtc2}.
This can be efficiently parallelized using graphics processing units (GPUs)~\citep{Talbot2019} to control the run time. However, this method fails for very narrow distributions, due to having only limited samples from the posterior.

In this work, we consider the converse weighting for this marginalization step: we sample from the population model and compute the likelihood of the observed data for each event given these samples.
This requires an efficient method for evaluating the likelihood at arbitrary points in parameter space.
Previous work has used Kernel Density Estimates (KDEs, \cite{Rosenblatt1956}) and Gaussian Processes (GPs, \cite{Rasmussen2006}) for density estimation \citep{Wysocki2020, Landry2019, Demilio2021}. While KDEs can be made quickly, distributions with sharp edges and high dimensions can cause the KDE to break down and the complexity of evaluating the KDE scales with the number of samples in the distribution. Similarly, GPs have been shown to provide good fits in small dimensions, but finite-binning effects from fitting the histogrammed samples can limit the accuracy of the density estimate. Another GP method for making density estimates is used in the parameter estimation code \textsc{RIFT} \citep{rift}. However, this requires fixed sets of intrinsic parameters and can be unsuitable for analyzing relationships between source-frame parameters, which is needed in this work (see Section \ref{Models}). 

While these methods provide estimates of single-event likelihoods, they each involve assumptions and/or computational complexities which may make them sub-optimal for any general given hierarchical inference problem \citep{Demilio2021, talbot2020}. In the next section, we make density estimates of single-event likelihoods using GMMs for use in a hierarchical inference framework. The steps presented here provide a relatively computationally-inexpensive density estimation procedure that avoids the shortcomings of the methods outlined above. 

\subsection{Density Estimation}

We begin with the goal of being able to evaluate the individual likelihoods $\mathcal{L}(d|\theta)$ at any arbitrary point in parameter space.
To do this we must begin with our $N$ sets of posterior samples (for $N$ events) and create a functional form for each likelihood.

As an estimate of the $D$-dimensional marginalized likelihood for an observed event, we model the likelihood as a linear combination of several $D$-dimensional Gaussians, where $D$ is the number of parameters of interest in each event's posterior.
In such a Gaussian mixture model (GMM), a density estimate is made from the set of discrete samples from the single-event posterior, resulting in an analytic model for the likelihood, allowing for evaluation of $p(d|\theta)$ for any $\theta$. 

\begin{figure}

    \includegraphics[width=\linewidth]{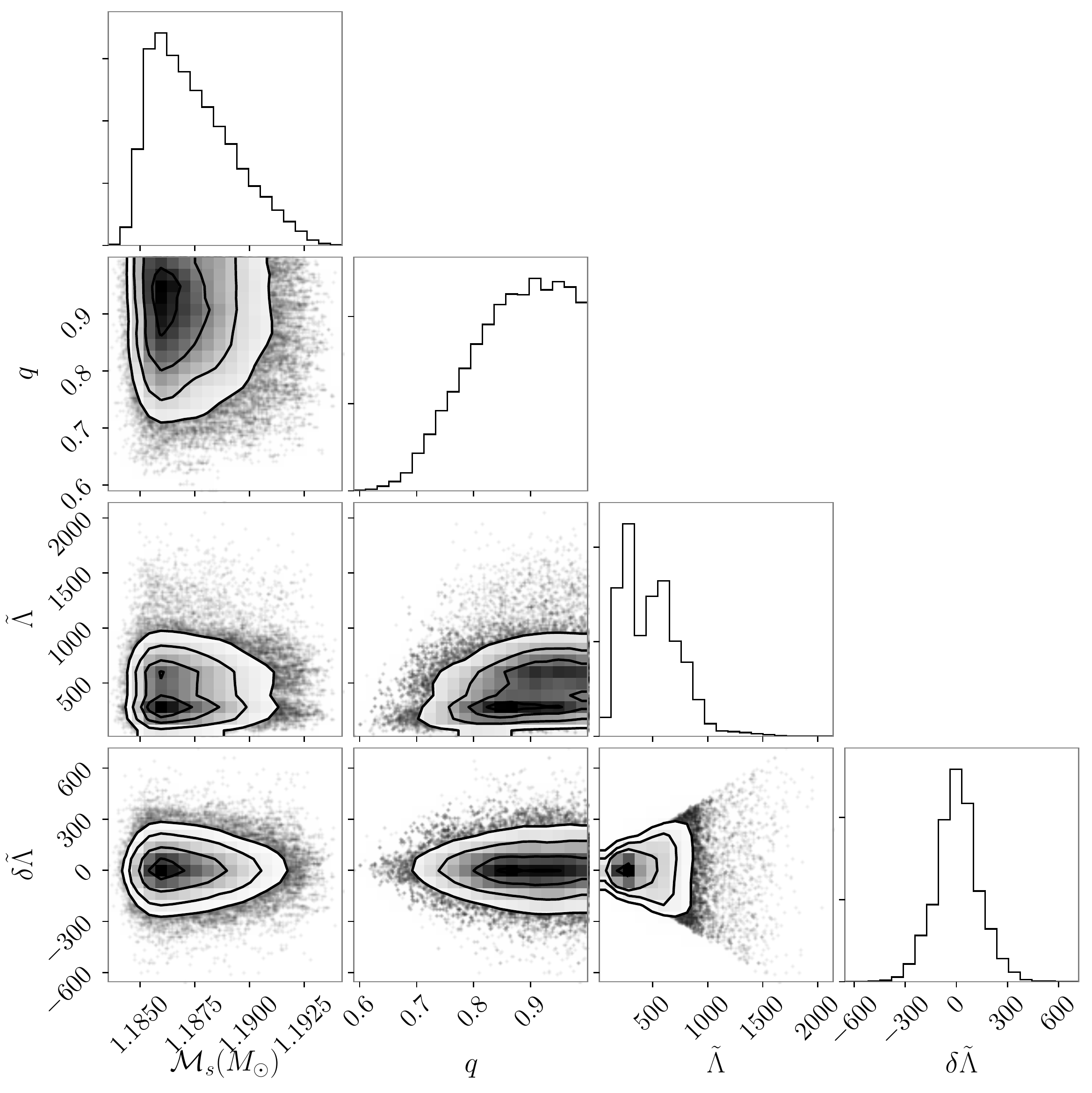}
    \includegraphics[width=\linewidth]{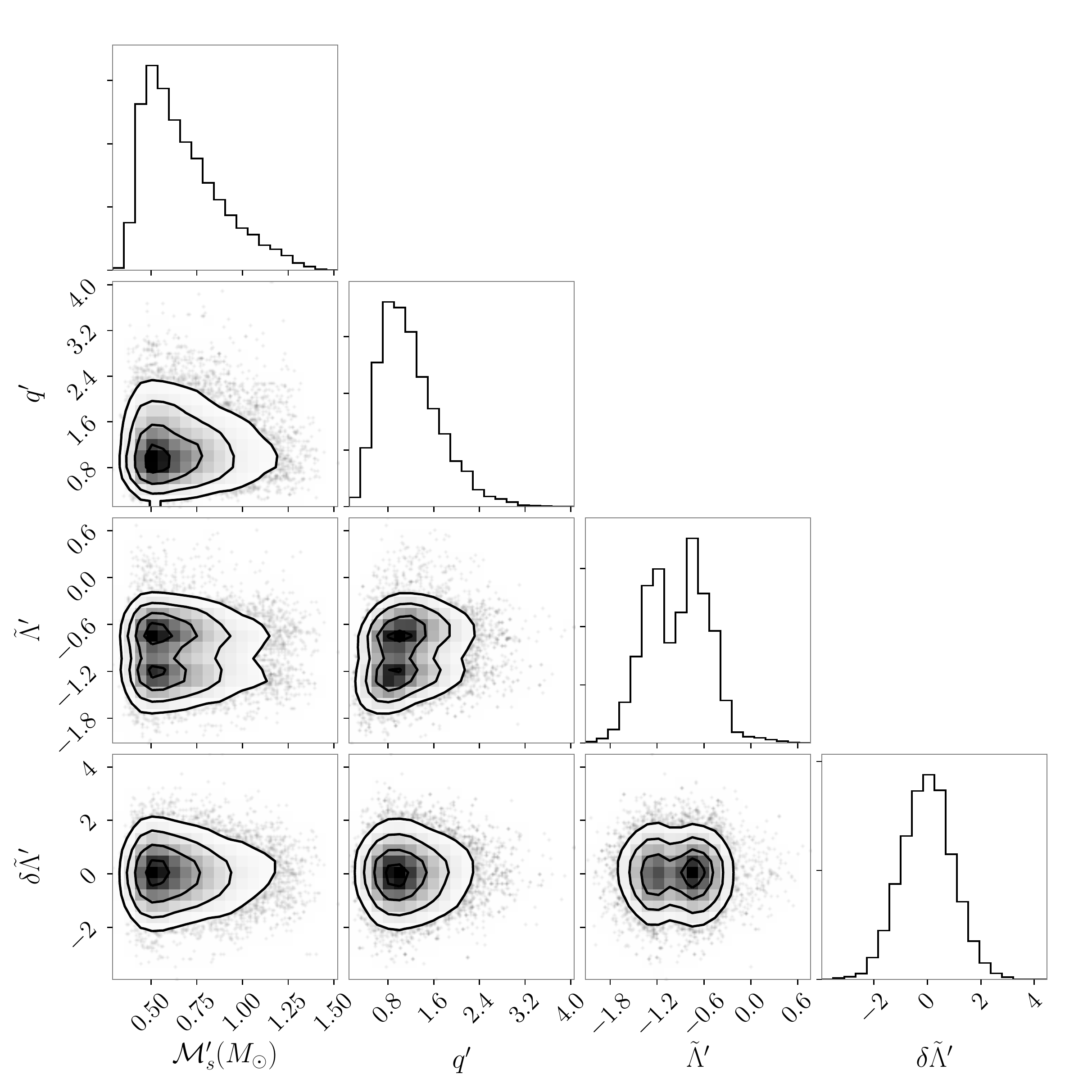}
    \caption{(Upper panel) Posterior distribution for the mass and tidal parameters of GW170817 in physical space (before transformation). (Lower panel) Posterior distribution for GW170817 after transforming samples into fitting space. Note that the domain of the transformed samples is much more uniform across parameter space and the sharp edges both in 1-D and 2-D histograms have been removed.}
    \label{fig:170817}
\end{figure}

\subsubsection{Pre-processing}

Realistic posterior distributions for parameters in gravitational wave analysis are subject to sharp edge effects, widely differing domains between parameters, and other features which make the raw distribution unsuitable for reliable fitting with our density estimation method. In the top panel of Figure \ref{fig:170817} we plot the posterior distribution for the chirp mass ($\mathcal{M}$), mass ratio ($q$), and two tidal parameters $\tilde{\Lambda}$, and $\delta\tilde{\Lambda}$ (see Section \ref{Data} for descriptions of these parameters) for GW170817 \citep{eos170817}. For this event, we note the marginal distribution for mass ratio $q$ exhibits a hard cutoff at $q=1$, and the correlation plot between $\tilde{\Lambda}$ and $\delta\tilde{\Lambda}$ has a sharp triangular shape.

We follow \cite{talbot2020} to map the observed distribution to one that is smoother and on a better-behaved domain, and use this transformed distribution to train the density estimate.

First, we map each posterior sample in $\theta$ to the unit interval using the cumulative distribution function (CDF) $F$ of the prior.
Next, we transform the samples from the unit interval to the unit normal distribution.
The transformed sample is 
\begin{equation} \label{transform}
    \theta' = \Phi^{-1}(F({\theta})),
\end{equation}
where $\Phi^{-1}$ is the probit function, the inverse CDF of the unit normal distribution \citep{Bliss1934}
\footnote{Note that if an analytic form of $F({\theta})$ is not known, an interpolant may be necessary to compute $F({\theta})$ for arbitrary values of $\theta$.}.

The result of this transformation on the posterior distribution for GW170817 can be seen in the lower panel of Figure \ref{fig:170817}.
The original posterior distribution (in physical space) can be compared to the transformed distribution, which has been made more suitable for fitting to a GMM.

\subsubsection{Fitting the Distribution}

After mapping the samples from the posterior distribution using the method in the previous section, the transformed samples follow a distribution more suitable to fitting with a GMM.

We train the model on the posterior samples and determine the maximum likelihood means, covariances, and weights assigned to each component of the GMM. Mathematically, the density estimate of an observed posterior distribution is
\begin{equation}\label{GMM}
    p(\theta'_i|d_i) 
    \approx \sum^K_{k} \phi_{ik} \mathcal{N}(\theta'_i|\mu_{ik}, \Sigma_{ik}) ,
\end{equation}
where the $k^{th}$ component of the mixture is a multivariate Gaussian of mean $\mu_k$ and covariance $\Sigma_k$, weighted by $\phi_k$. Here, $\sum^K_k \phi_{ik} = 1$ for a $K$-component GMM, for each $i$. Note that the index $k$ here runs over the \textit{components} (individual Gaussians) in the mixture model, not the observed events, as this mixture model is unique to each event.
We use the Gaussian mixture model as implemented in \textsc{Scikit-Learn} \citep{scikit-learn}, which uses an expectation-maximization method to fit for the $\mu$, $\Sigma$, and $\phi$ parameters in Equation \ref{GMM} conditioned on a set of transformed samples from the posterior distribution of the $i$th event. 

Since a GMM is a sum of individual weighted Gaussian components, we must determine how many such components to use to make the optimal fit characterizing the distribution without overfitting.
To determine this, we take a set of posterior samples from an event and we randomly assign 80\% of the transformed posterior samples to a training set and the other 20\% to a testing set.
We train the $K$-component GMM using the training data and evaluate the score (sample-wise average log-likelihood) of the testing samples for the GMM.
To determine the optimal number of components to use, we vary $K$ and repeat this process until the score noticeably flattens, indicating an increased value of $K$ does not better characterize the distribution.
When working with a catalog of observed events, it may be efficient to do this step using one selected event (possibly corresponding to the most complex posterior distribution), and use this optimal $K$ for all GMM density estimates in the catalog. However, a more complete fitting method would consist of fitting the for the optimal $K_i$ for each event $i$, rather than using the same $K$ for all events. We note that one could also fit for the optimal $K_i$ values for each event using a statistic such as the Bayesian information criterion (BIC) \citep{Kass95}. However, we find that the fits we obtain from the flattening of the score are sufficient for good recovery of our simulated distribution, as reported in Section \ref{Results}.

\begin{figure}
    \centering
    \includegraphics[width=\linewidth]{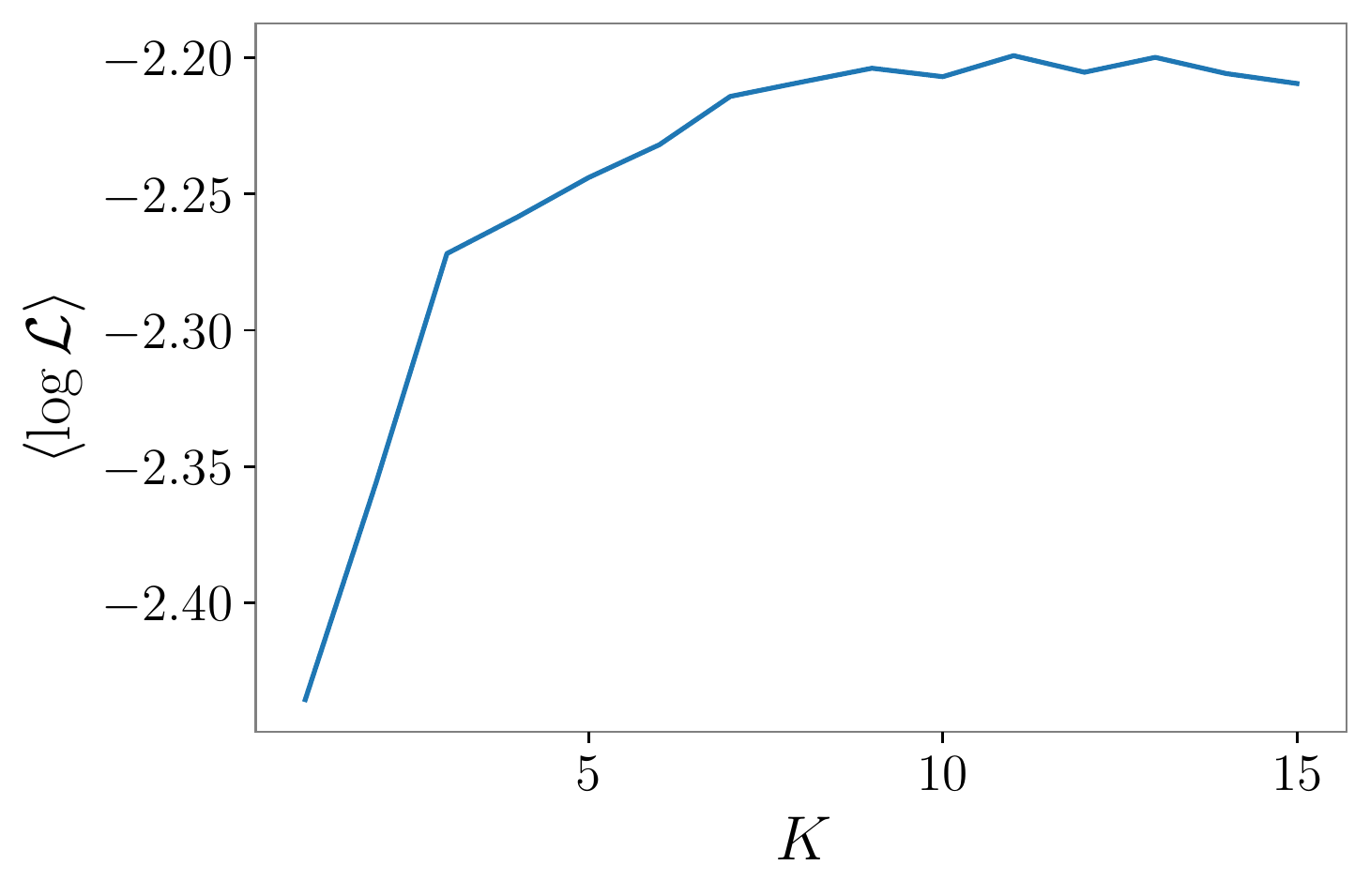}
    \caption{Average (natural) log likelihood of evaluation samples as a function of number of components ($K$) used to generate GMM, using posterior samples from GW170817. The score flattening out by $K \approx$ 8-10 indicates that the GMM does not provide a better density estimate for larger $K$.}
    \label{fig:scores}
\end{figure}

\begin{figure}
    \centering
    \includegraphics[scale=0.35]{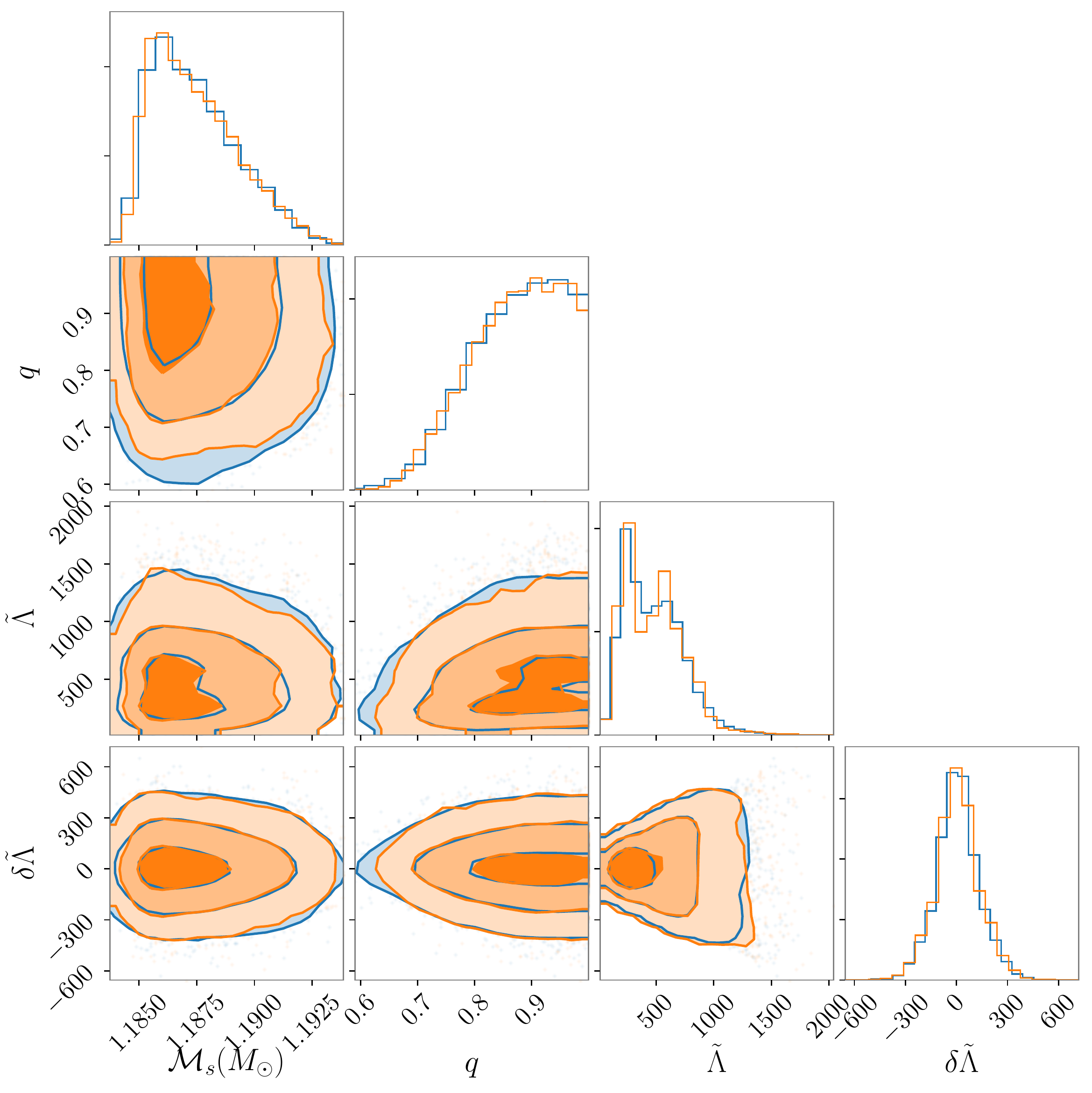}
    \centering
    \includegraphics[scale=0.35]{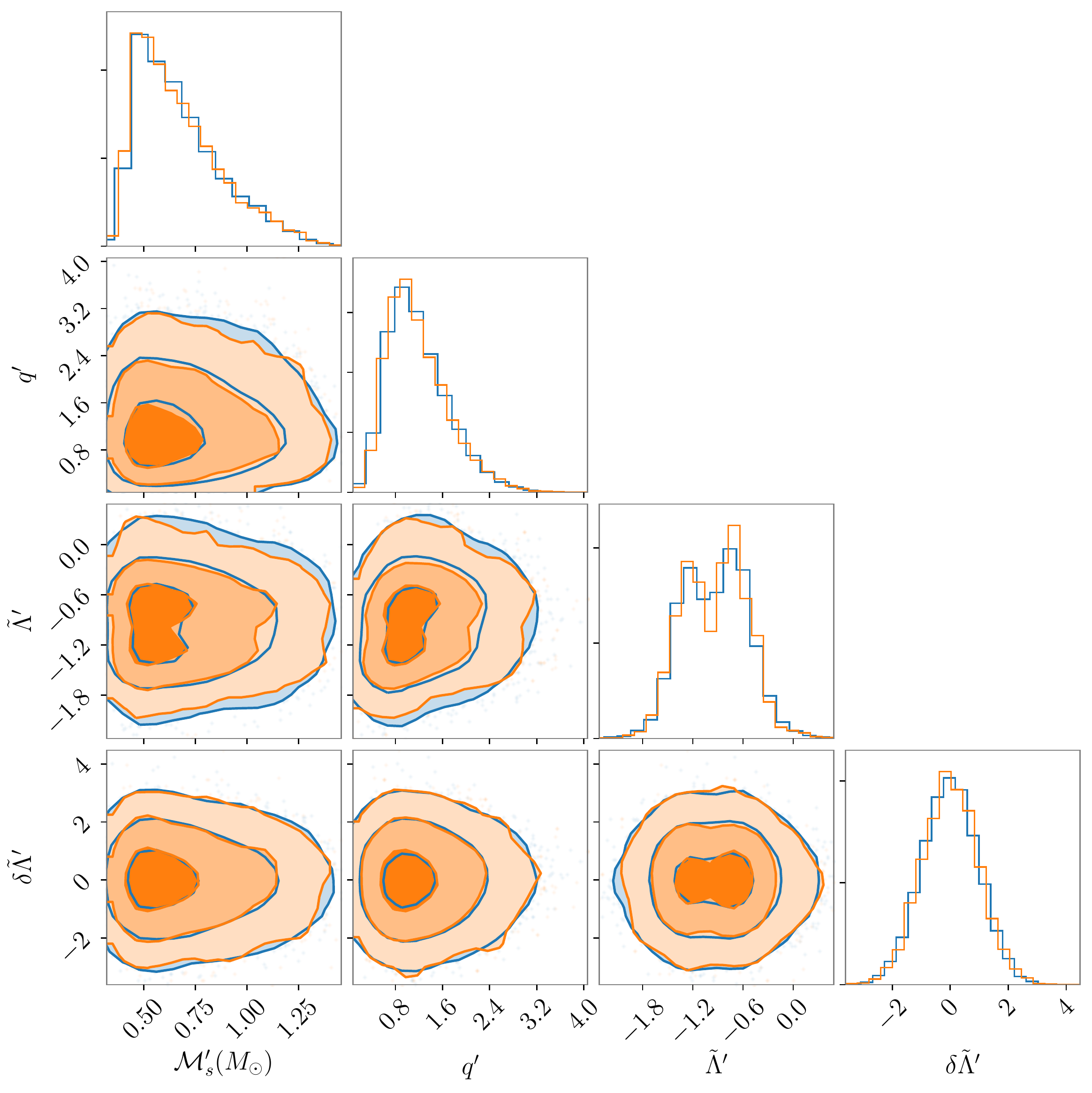}
    \caption{4-dimensional posterior distribution for GW170817. Orange is the posterior samples and blue is from samples of the GMM fit. The overlap between the two distributions shows the GMM provides a good density estimate. Plotted in physical (transformed) space in the upper (lower) panel.
    }
    \label{fig:gmm}
\end{figure}

In Figure \ref{fig:scores}, we show this curve using our GMM fits from GW170817.
As the score flattens out by $K = 8-10$ components, this represents the optimal number of components to use in the GMM for this posterior distribution. As an illustration, we compare the GMM fit to the true posterior distribution using samples drawn from the GMM and the original transformed samples for GW170817 in Figure \ref{fig:gmm}.

Since the GMM is trained on transformed \textit{posterior} samples, we convert the GMM density estimate $\mathcal{D}_i(\theta')$ from the $i$th event into a single-event \textit{likelihood} via:
\begin{equation}\label{likelihoodgmm}
    \mathcal{L}_i(\theta ') = \frac{\mathcal{D}_i(\theta ')}{\mathcal{N}(\theta ', \mu = 0, \sigma = 1)}.
\end{equation}
This results in the correct likelihood because we used the sampling priors for $F(\theta)$ in the original transformation $\theta \rightarrow \theta '$ (see Equation \ref{transform}).

\subsection{Hierarchical Likelihoods using Density Estimates}

For a given population model $p(\theta|\Omega)$, we compute the likelihood of the event $i$ (for $i \in (1, N)$) as:
\begin{equation}
    \mathcal{L}_i(d_i|\Omega) = \frac{1}{M} \sum_{\theta \sim p(\theta|\Omega)} \mathcal{L}_i(\theta') ,
\end{equation}
where $\mathcal{L}_i$ is the likelihood of the $i$th event (Equation \ref{likelihoodgmm}), and $M$ samples $\theta$ are drawn from the population model $p(\theta|\Omega)$.
This is a practical Monte Carlo integration scheme for the integral in Equation \ref{hyperlikelihood}, dependent on the ability to sample from the population model and evaluate the single-event likelihoods at the corresponding points in parameter space.

An implicit step in the above hierarchical likelihood equation is the mapping of the population samples $\theta$ into the corresponding transformed (fitting) space samples $\theta'$ for each density estimate, to match the space of the density estimates. 

The total likelihood (i.e. Equation \ref{hyperlikelihood}) for the data from $N$ events therefore becomes:
\begin{equation}\label{MCLike}
    \mathcal{L}(\mathbf{d}|\Omega) = \prod_{i=1}^{N}\frac{1}{M} \sum_{\theta \sim p(\theta|\Omega)} \mathcal{L}_i(\theta').
\end{equation}
Although \textsc{Scikit-Learn} provides a method of computing the log-likelihood of samples in a GMM fit, for a single evaluation of the total likelihood, each of $N$ GMMs must be evaluated for $M$ samples, which can become a computational burden for large $N$ and $M$.
When there are many observed events, it becomes more efficient to extract the best-fit means, covariances, and weights from the GMM fits and evaluate the likelihood matrices in Equation \ref{GMM} directly on a GPU using \textsc{CuPy} \citep{cupy}.
This avoids explicitly looping over the $N$ GMMs for each evaluation of the joint likelihood, while efficiently performing computations over the $\mathcal{O}(N \times M \times K)$ array using array broadcasting and vectorization on a GPU. 

\section{Models}\label{Models}

For the implementation of our GMM-based hierarchical inference method, we let $\Omega$ characterize the mass distribution as well as the EOS relating the $\Lambda$ and $m$ parameters. This requires choosing parameterized models for the mass distribution and the EOS model.

\subsection{Mass Population Model}

To model the distribution of neutron stars in merging binaries, we consider the observationally-motivated framework in \cite{farrow2019}.
In that work, the authors found evidence based on observations of galactic BNS systems for each companion of a BNS system being drawn from a separate population distribution.
The first population distribution characterizes the member of the binary that forms first and spins up due to accretion, known as a recycled neutron star.
The other member of the system is known as the slow neutron star, as it is born second and spins down quickly after formation.
The authors found the model with the best support consists of a two-component Gaussian model for the recycled neutron star mass distribution, and a uniform distribution for the slow neutron star mass.

Adopting this mass distribution model, the subset of $\Omega$ describing the mass population consists of 8 parameters.
The lower-mass Gaussian of the recycled distribution is described by the parameters $(\mu_{R,a}, \sigma_{R,a})$ and is weighted by $\alpha$, and the higher-mass component is described by $(\mu_{R,b}, \sigma_{R,b})$. The probability of observing a mass $m$ from the recycled mass distribution is
\begin{equation}\label{recycled}
\begin{split}
    p_R(m|\Omega) & = \alpha \mathcal{N}(m| \mu_{R,a}, \sigma_{R,a}) \\
                & + (1 - \alpha) \mathcal{N}(m|\mu_{R,b}, \sigma_{R,b}).
\end{split}
\end{equation}
For the slow mass distribution, we denote the low and high limits as $m_{S,l}$ and $m_{S,h}$, respectively. The maximum mass parameter, $M_{\max}$ represents an absolute cutoff of both distributions; we truncate the recycled (Equation \ref{recycled}) and slow mass distributions at $M_{\max}$ on the high end and 1 $M_{\odot}$ on the low end.

By considering the observed galactic BNS systems in the context of binary formation and evolution models, \cite{zhu2020} found that modeling the slow companion as non-spinning was a robust approximation for gravitational wave data analysis. The recycled partner, while spun-up from the slow companion, has very little support for spins of $\chi > 0.05$ for population and EOS models they considered. We therefore do not consider spins at all in this work, and model all sources as nonspinning. Since we neglect spins, we have no way of concretely knowing which component mass $(m_1, m_2)$ represents the slow or recycled mass, so each computation of the population probability must account for the possibility of either component being drawn from either distribution, with the constraint that each BNS system consists of exactly one recycled and one slow neutron star.

\subsection{EOS Model: The $\Lambda$--m Relation}

Several nonparametric and parametric models for EOS-sensitive observables exist in gravitational wave literature, based on the assumption that a neutron star of a given mass will have a corresponding $\Lambda$ uniquely determined by its EOS \citep{Read2009, Lindbloom2010, Landry2019}. Therefore, recovering parameters characterizing this mapping between the two observables, $m$ and $\Lambda$, may provide a way to recover information about the underlying nuclear EOS.

To model the EOS-sensitive relationship, we follow the examples of \cite{Agathos2015, delpozzo2013} and consider a simple expansion of $\lambda(m)$ about the canonical reference value of 1.4 $M_\odot$:

\begin{equation}\label{lambda}
    \lambda(m|c_0, c_1) = c_0 + c_1 \left(\frac{m - 1.4 M_\odot}{M_\odot}\right)
\end{equation}

The expansion coefficients in this model are our EOS-sensitive parameters, with different combinations approximating different EOSs. Previous work has shown that with this parameterization, LIGO observations will be unlikely to resolve terms $c_j$ for $j \gtrsim 1$, so we only include these first two terms in our demonstration. With component masses of observed galactic BNS systems peaking around 1.4 $M_\odot$, this form for $\lambda(m)$ can provide meaningful constraints on the EOS, as the expansion is centered at this value. 

\begin{figure}
    \centering
    \includegraphics[width=\columnwidth]{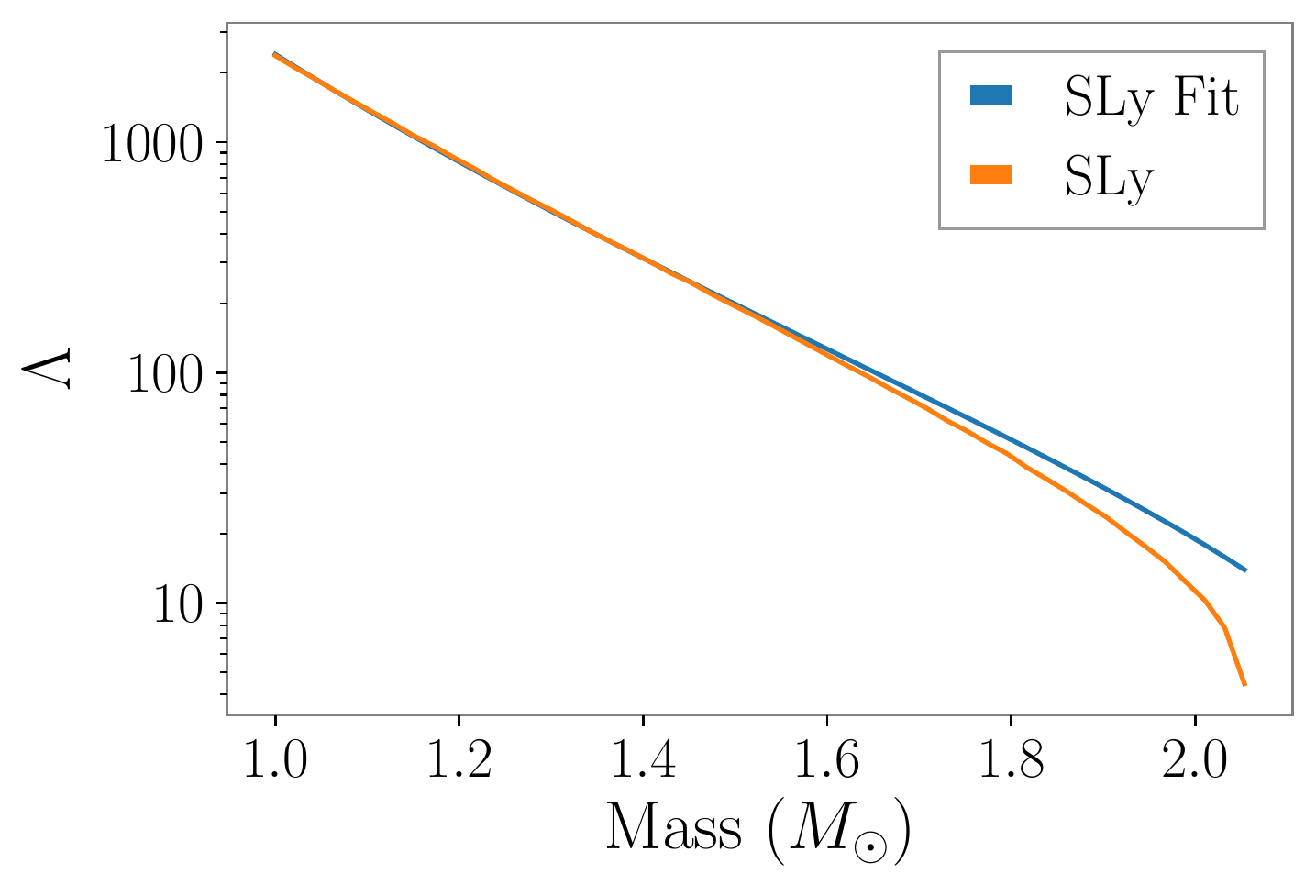}
    \includegraphics[width=\columnwidth]{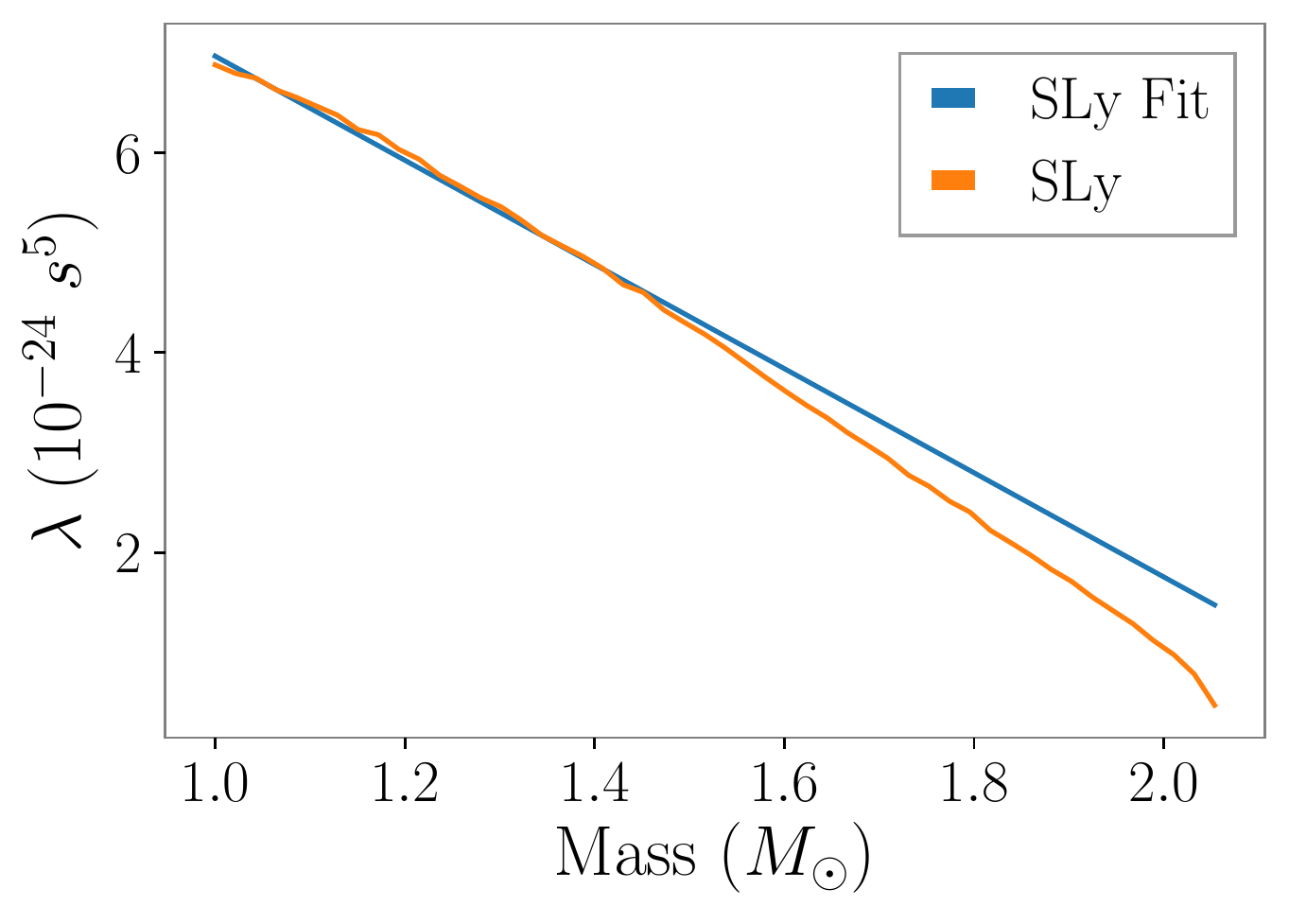}
    \caption{Dimensionless tidal deformability (upper panel) and tidal deformability (lower panel) as a function of mass. Units of $\lambda$ are in seconds to the fifth power, following the convention in \cite{Agathos2015}.}
    \label{fig:sly}
\end{figure}

For our fiducial choice of EOS to simulate, we use the values of $c_0$ and $c_1$ corresponding to the relatively-soft SLy EOS \citep{Douchin2001}. In Figure~\ref{fig:sly}, we show the linear fit to the true EOS from Equation~\ref{lambda} for the $\Lambda-m$ (top) and $\lambda-m$ (bottom) relationships.
The approximation breaks down for $m > 1.8 M_\odot$ as previously noted in, e.g,~\cite{Chatziioannou2020}, however, the majority of simulated events considered in this work are less massive than this.
If observed BNS mergers contain high mass components, this linear parameterization will not be valid; however, it is sufficient for our proof-of-principle (see Section \ref{Discussion}).

This model provides a weak connection between our EOS-sensitive model and the mass population model. Since $c_0$ and $c_1$ are allowed to vary in Equation \ref{lambda}, it is possible to arrive at a negative value for $\Lambda$, which is unphysical. Since $\Lambda(m)$ is a decreasing function with mass, we therefore constrain $M_{\max}$ such that $0 < \Lambda(M_{\max}) < 200$, consistent with limits on minimum $\Lambda(m)$ for commonly-used EOSs (see Figure 1 in \cite{Chatziioannou2020}).

It is worth noting the limited significance of the $M_{\max}$ parameter in this work, as the constraint $0 < \Lambda(M_{\max}) < 200$ is simply a cutoff to keep values of $\Lambda$ physical.
This does not necessarily correspond to the maximum neutron star mass as determined by the EOS; this is known as the Tolman-Oppenheimer-Volkoff mass ($M_{\textrm{TOV}}$), and is determined from the stability conditions set by a particular EOS \citep{Kalogera1996}.
By cutting off the population model at $M_{\max}$ in this work, we do not associate $M_{\max}$ with the most massive \textit{possible} neutron star, but instead $M_{\max}$ is the upper limit on the mass of neutron stars in merging binaries.
Thus, stellar binary evolution and population channel models play an additional role in the significance of $M_{\max}$ in the population.
A more rigorous approach may be to calculate $M_{\mathrm{TOV}}$ for a given EOS and enforce the condition $M_{\max} < M_{\mathrm{TOV}}$.

Although this work is a proof-of-concept for this method, improvements to the EOS modeling and parameterization can make the results more realistic and applicable to real observations.
The breakdown of our choice of EOS model at high masses serves to potentially bias the entire inference if a significant number of events are observed far from the reference mass used in the Taylor expansion of Equation \ref{lambda}.
Additionally, previous work has shown that a simple Taylor expansion, such as the model used in this work, may not be robust to EOS models with phase transitions to non-hadronic constituents.
Therefore, a more general model robust to arbitrary EOSs without deviations at high masses may be useful for realistic observations. 

\section{Data and Implementation}\label{Data}

\subsection{Data}
\begin{table}[b]
\centering
 \begin{tabular}{||c c c c||}
 \hline
 Parameter & Value & Prior & Units\\ [0.5ex] 
 \hline\hline
 $\mu_{Ra}$ & 1.34 & (1, 2)  & $M_\odot$\\ 
 \hline
 $\sigma_{Ra}$ & 0.02 & (0.005, 0.5) & $M_\odot$ \\
 \hline
 $\mu_{Rb}$ & 1.47 & ($\mu_{Ra}$, 2) & $M_\odot$ \\
 \hline
 $\sigma_{Rb}$ & 0.15 & (0.005, 0.5) & $M_\odot$ \\
 \hline
 $\alpha$ & 0.68 & (0, 1) & N/A \\ 
 \hline
 $M_{Sl}$ & 1.16 & (1, 1.7) & $M_\odot$ \\
 \hline
 $M_{Sh}$ & 1.42 & ($M_{Sl}$, $M_{\max}$) & $M_\odot$ \\
 \hline
 $M_{\max}$ & 2.2 & (1.9, 2.3) & $M_\odot$\\ 
 \hline
 $c_0 / 10^{-24}$ & 4.88 & ($10^{-1}$, 10)  & $s^5$ \\
 \hline
 $c_1 / 10^{-24}$ & -5.21 & (-10, -1) & $s^5$ \\[1ex] 
 \hline
\end{tabular}
\caption{Summary of hyperparameters used in demonstration. Value column indicates the number used to generate the samples, and the prior column is the bounds of the uniform prior used for sampling. The mass distribution parameters include the means, $\mu_R$, and standard deviations, $\sigma_R$ of the low-mass Gaussian, a, and higher-mass Gaussian, b, of the recycled mass distribution, along with a weight $\alpha$ (b weighted by $(1 - \alpha)$). The slow mass uniform distribution is characterized by its lower bound $M_{Sl}$ and upper bound $M_{Sh}$. See the dashed line in Figure \ref{fig:ppd} for probability density function (PDF) of the input mass distribution. The $c_0$ and $c_1$ parameters are the EOS-sensitive parameters in Equation \ref{lambda}. All mass parameters in units of $M_\odot$.}
\label{tab:params}
\end{table}

In order to demonstrate our method, we simulate 100 BNS signals drawn from the mass distribution characterized by the maximum likelihood estimate in \cite{farrow2019}. This corresponds to the combination of mass distribution parameters listed in Table \ref{tab:params}. We specify injected tidal parameters using our linear fit to the SLy EOS, neglecting spins.

We draw the extrinsic parameters isotropically in position and orientation with distances uniform in source frame between $20-300$ Mpc using the cosmology from the \textsc{Planck} 2015 data release~\citep{planck2015}. 
\\
For each simulated signal, we generate $128s$ of colored Gaussian noise corresponding to the three-detector Advanced LIGO-Virgo network operating at their projected design sensitivities~\citep{observing_prospects}.
We employ a GPU-implementation of the \textsc{TaylorF2} waveform model~\citep{Buonanno09, Talbot2019} and analyze data between $20-2048$ Hz. We simplify our parameter estimation by marginalizing phase, merger time, and luminosity distance for sampling the posterior distribution for single-event analyses. We reconstruct the luminosity distance marginal posterior distribution in post-processing using the method outlined in \cite{Thrane:2018qnx}.

We impose a detection threshold of $\text{SNR} > 8$ in at least one detector.
For the 37 events passing our detection threshold, we infer the posterior distribution using the \textsc{Bilby}~\citep{bilby} implementation of \textsc{PyMultiNest}~\citep{pymultlinest}.
\\
We employ a uniform prior on detector-frame chirp mass ($\mathcal{M}_d$) of $\pm 0.01 M_\odot$ around the injected value for each event. Our prior on $q$ is uniform from 0.125 to 1. For $\tilde{\Lambda}$, defined as \citep{wade2014}
\begin{equation}\label{lambdatilde}
\begin{split}
    \tilde{\Lambda} = \frac{8}{13} &\Big[(1 + 7\eta - 31\eta^2)(\Lambda_1 + \Lambda_2) \\ &+ \sqrt{1 - 4\eta} (1 + 9\eta - 11\eta^2)(\Lambda_1 - \Lambda_2) \Big],
\end{split}
\end{equation}
we use a uniform prior from 0 to 5000. Here, $\eta \equiv q (1 + q)^{-2}$ is the symmetric mass ratio. We construct a conditional sampling prior on $\delta\tilde{\Lambda}$ as follows: for each sample $\theta_i$, we analytically compute the maximum and minimum allowed values of $\delta{\tilde{\Lambda}}_i$ conditioned on $q_i$ and $\Lambda_i$. The parameter $\delta\tilde{\Lambda}$ is defined as \citep{wade2014}
\begin{equation}
\begin{split}
    \delta\tilde{\Lambda}(q, \tilde{\Lambda}) & = \frac{1}{2}\Big[\sqrt{1 - 4\eta} \left(1 - \frac{13272}{1319}\eta + \frac{8944}{1319}\eta^2\right)(\Lambda_1 + \Lambda_2) \\ + 
    & \left(1 - \frac{15910}{1319}\eta + \frac{32850}{1319}\eta^2 + \frac{3380}{1319}\eta^3\right) (\Lambda_1-\Lambda_2) \Big],
\end{split}
\end{equation}
such that $\delta\tilde{\Lambda}$ deviates from 0 as the differences in component tidal deformabilities increases. It therefore reaches a maximum (minimum) when $\Lambda_1$ ($\Lambda_2$) is 0. 
Thus, we calculate the bounds of the uniform prior on $\delta\tilde{\Lambda}$ conditioned on a sample of $(q, \tilde{\Lambda})$, where $\Lambda_{1 (2)}$ is computed by setting $\Lambda_{2 (1)} = 0$ for fixed $q$ and $\tilde{\Lambda}$ in Equation \ref{lambdatilde}, and using the resulting value of $\delta\tilde{\Lambda}$ as the upper (lower) bound.
We then consider a uniform prior on $\delta\tilde{\Lambda}$ from these conditions. 

While $\mathcal{M}$ is well-constrained in the detector frame, the hyperparameters we consider in this work are only relevant to source-frame masses.
Therefore, if a given set of posterior samples only contain $\mathcal{M}$ in the detector frame, they must be converted to source frame via the relationship $\mathcal{M}_{d} = (1 + z) \mathcal{M}$.
We construct our corresponding prior on $\mathcal{M}_s$ as \citep{Thrane:2018qnx},
\begin{equation}
    \pi(\mathcal{M}_s) = \int dz d\mathcal{M}_d \pi(z) \pi(\mathcal{M}_d) (1+z) \delta\left(\mathcal{M}_s - \frac{\mathcal{M}_d}{(1+z)}\right).
\end{equation}
This is marginalized over both detector-frame mass and distance.
This relation may be unique to each observed event if a different prior on $\mathcal{M}_d$ is used for each event.
We therefore associate a unique prior on $\mathcal{M}$ with each GMM \footnote{Note we express $\mathcal{M}_s$ = $\mathcal{M}$ in this work, with detector-frame chirp mass written explicitly as $\mathcal{M}_d$}. 

To make the density estimates of each posterior distribution, we follow the method outlined in the previous section, using our single-event sampling priors for $F$ in Equation \ref{transform} for the transformation into fitting space for each event's posterior. 
As an example of GMM density estimation on BNS posterior distributions, we show the fit to GW170817 in Figure \ref{fig:170817}.
We observe that BNS posteriors may include strong correlations between $\tilde{\Lambda}$ and $\delta{\tilde{\Lambda}}$ (i.e. the triangular shape in the $\tilde{\Lambda} - \delta{\tilde{\Lambda}}$ correlation plot in Figure \ref{fig:170817}), possibly impacting the quality of the density estimate.
Using the CDF of our conditional prior as $F(\theta)$, the transformation decorrelates $\tilde{\Lambda}$ and $\delta\tilde{\Lambda}$.
As can be seen in the bottom panel of Figure \ref{fig:170817}, the correlation between the tidal parameters no longer exists in transformed space when imposing this condition on the sampling prior.

To motivate our choice for the number of components in our GMMs, we implement the scoring method described in Section \ref{Methods} on GMM density estimates of GW170817 and shown in Figure~\ref{fig:scores}.
Based on this example, we use 10 components for each GMM density estimate of our simulated BNS events.

\subsection{Sampling the Hyper-posterior}

For each calculation of the likelihood, we sample $M = 15,000$ masses from the population model and compute the corresponding tidal deformabilities conditioned on EOS-sensitive hyperparameters via the relationship in Equation \ref{lambda}.
We then convert the $M$ samples of $(m_r, m_s, \Lambda_r, \Lambda_s)$ to $(\mathcal{M}, q, \tilde{\Lambda}, \delta\tilde{\Lambda})$ and then into the fitting space for each of the GMMs. While we sample component masses in terms of recycled and slow mass, we convert $(m_r, m_s)$ to $(m_1, m_2)$, adopting the convention $m_1 > m_2$.
Using these transformed samples, we can evaluate Equation~\ref{MCLike}.

To calculate $P_{\text{det}}(\Omega)$ we draw 20,000 binaries from a mass distribution that is uniform in $[1, 2.2] M_{\odot}$.
For each simulated binary, we compute the SNR in an independent noise realization and keep those that pass our threshold.
We neglect the impact of tidal effects on sensitivity.
Since our mass distribution model is in terms of slow and recycled components, but our analysis can only specify $m_1$ and $m_2$, for the purposes of computing $p(\theta|\Omega)$ we assume \textit{a priori} each object is equally likely to be the recycled or slow companion, with the assumption each binary system contains exactly one slow and one recycled partner.

Each likelihood evaluation requires $\mathcal{O}(M \times N \times K) \sim 6 \times 10^6$ computations.
Running on an NVIDIA GeForce RTX 3080 GPU, each full likelihood evaluation took $\lesssim$ 50 ms for our 37 events which is comparable to the evaluation time for the method currently used to infer binary black hole mass distributions in LIGO-Virgo-KAGRA analyses~\citep{gwtc2, Talbot2019}.
We note that the sampling, transformation, and selection function steps in the likelihood introduce subdominant effects to computation time relative to the computation of $M \times N \times K$ Gaussian likelihoods.
We sample the hyper-posterior using the \textsc{Bilby} wrapper of the nested sampler \textsc{PyMultinest} \citep{pymultlinest}, sampling with 250 live points.

\section{Results}\label{Results}

\begin{figure}
    \centering
    \includegraphics[scale=0.6]{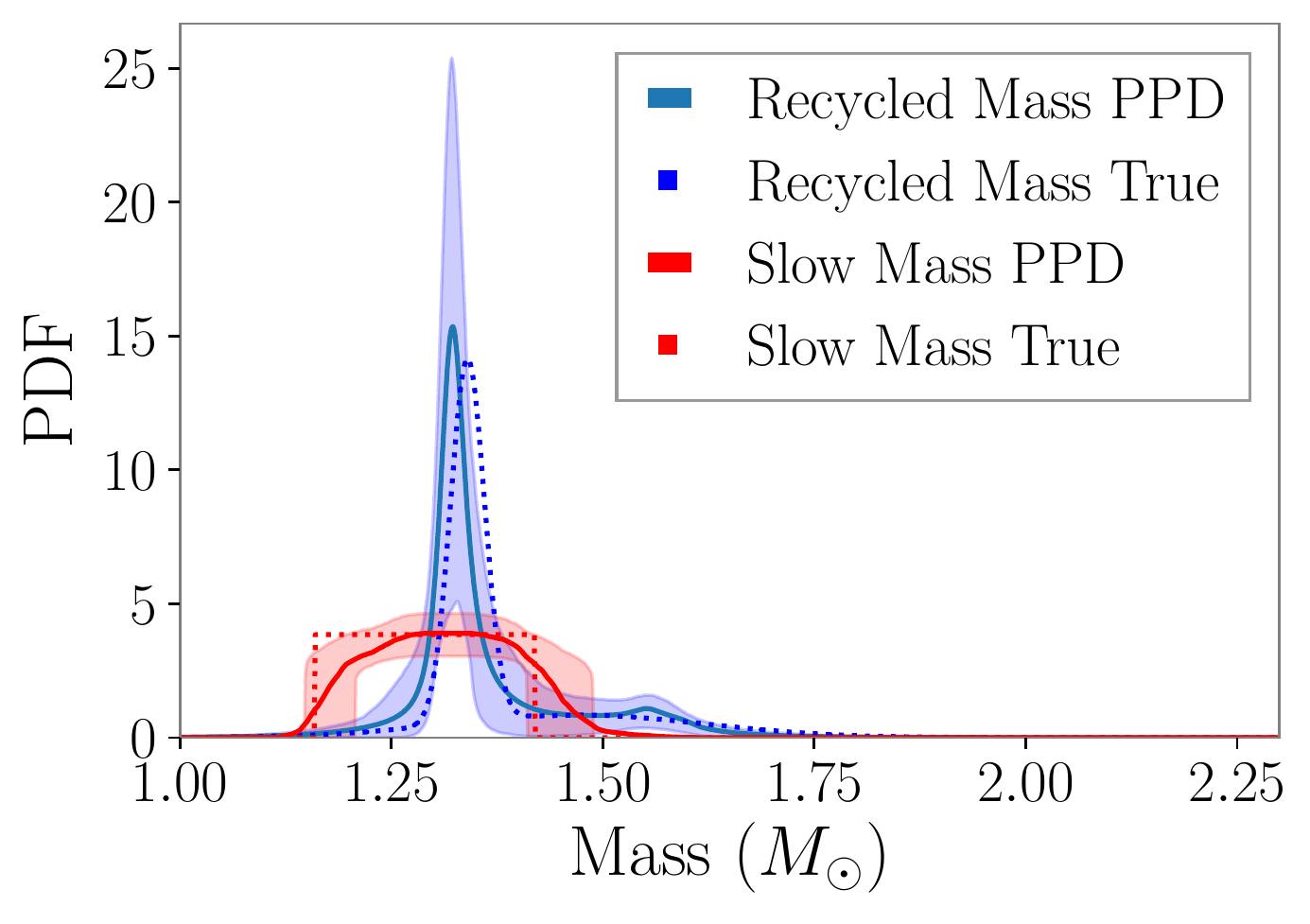}
    \caption{Inferred mass distribution from the full mass + EOS analysis of 37 simulated events. Solid lines represent the posterior predictive distribution (PPD). The recycled (slow) distribution is colored blue (red), with shading representing the $\pm 1\sigma$ $(68\%)$ credible region from the posterior. Dashed lines show the input distribution.}
    \label{fig:ppd}
\end{figure}

In Figure \ref{fig:ppd} we show the inferred mass distribution when the mass and EOS hyperparameters are sampled simultaneously.
The solid line shows the posterior predictive distribution (PPD), the shaded regions show the symmetric 68\% credible region, and the dashed lines show the true simulated distribution.
With the priors on the mass distribution hyperparameters spanning a wide range, the posterior distribution is relatively well-constrained around the input hyperparameters (see Figure \ref{fig:fullcorner}) for the full one- and two-dimensional posterior distributions.

Of note, we confidently recover the presence not only of the large peak in recycled mass distribution at $1.34 M_\odot$, but also the small and wide peak at higher masses, as shown by the $\mu_{Rb}$ and $\sigma_{Rb}$ panels in Figure \ref{fig:fullcorner}, as well as in the PPD in Figure \ref{fig:ppd}. The inferred location of the large peak is $\mu_{Ra} = 1.32^{+0.03}_{-0.03} M_\odot$ (all ranges are 90\% credible intervals), constrained to within $\sim$4\% of the input value. It is worth noting that the hyperparameters associated with the second peak of the recycled-mass distribution unsurprisingly show the poorest recovery.
This is expected as the second peak in this distribution is very small (i.e. most masses from the recycled distribution are in the lower-mass peak) and thus very few events coming from these masses are expected.
Nevertheless, we are able to recover evidence of this second peak around its input location. The presence of a secondary peak in the recycled mass distribution is favored over a single Gaussian, with a Savage-Dickey density ratio giving a Bayes factor of 2.6 in favor of a secondary peak ($\alpha \neq 0$ or $\alpha \neq 1$).

Additionally, the bounds of the slow mass distribution are well-constrained, with inferred values of $M_{Sh} = 1.45^{+0.08}_{-0.06} M_\odot$ and $M_{Sl} = 1.17^{+0.08}_{-0.04} M_\odot$. Both of these parameters are therefore constrained to within $\sim$ 10\% of their input values of $1.42 M_\odot$ and $1.16 M_\odot$, respectively.

\begin{figure}
    \centering
    \includegraphics[scale=0.59]{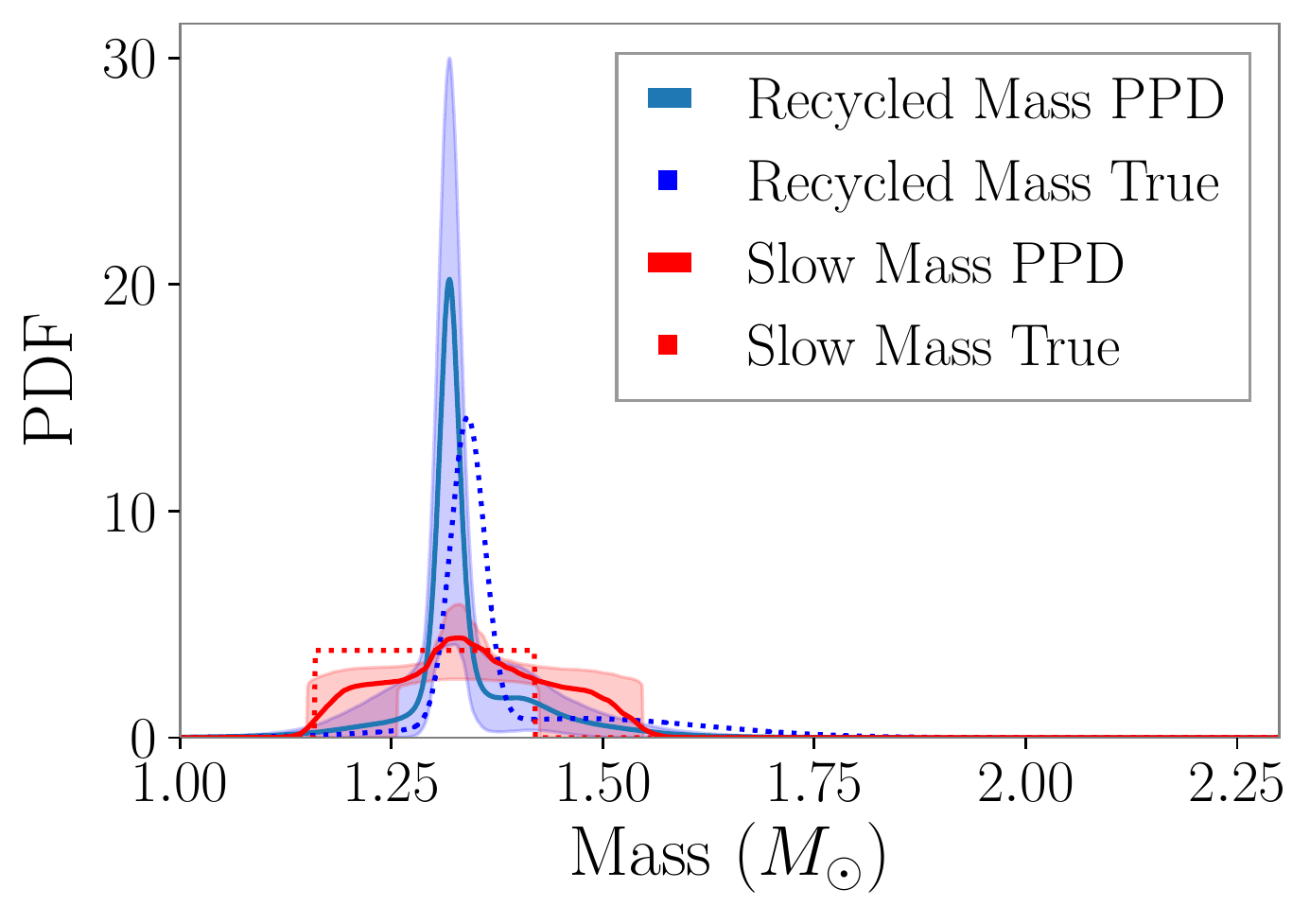}
    \caption{Inferred mass distribution from the mass-only analysis of 37 simulated events. Solid lines represent PPD. The recycled (slow) distribution is colored blue (red), with shading representing the $\pm 1\sigma$ $(68\%)$ credible region from the posterior. Dashed lines show the input distribution. Compare to Figure \ref{fig:ppd} to observe the bias in mass distribution recovery due to not including EOS inference.}
    \label{fig:massonlyppd}
\end{figure}

\begin{figure*}
    \includegraphics[width=\columnwidth]{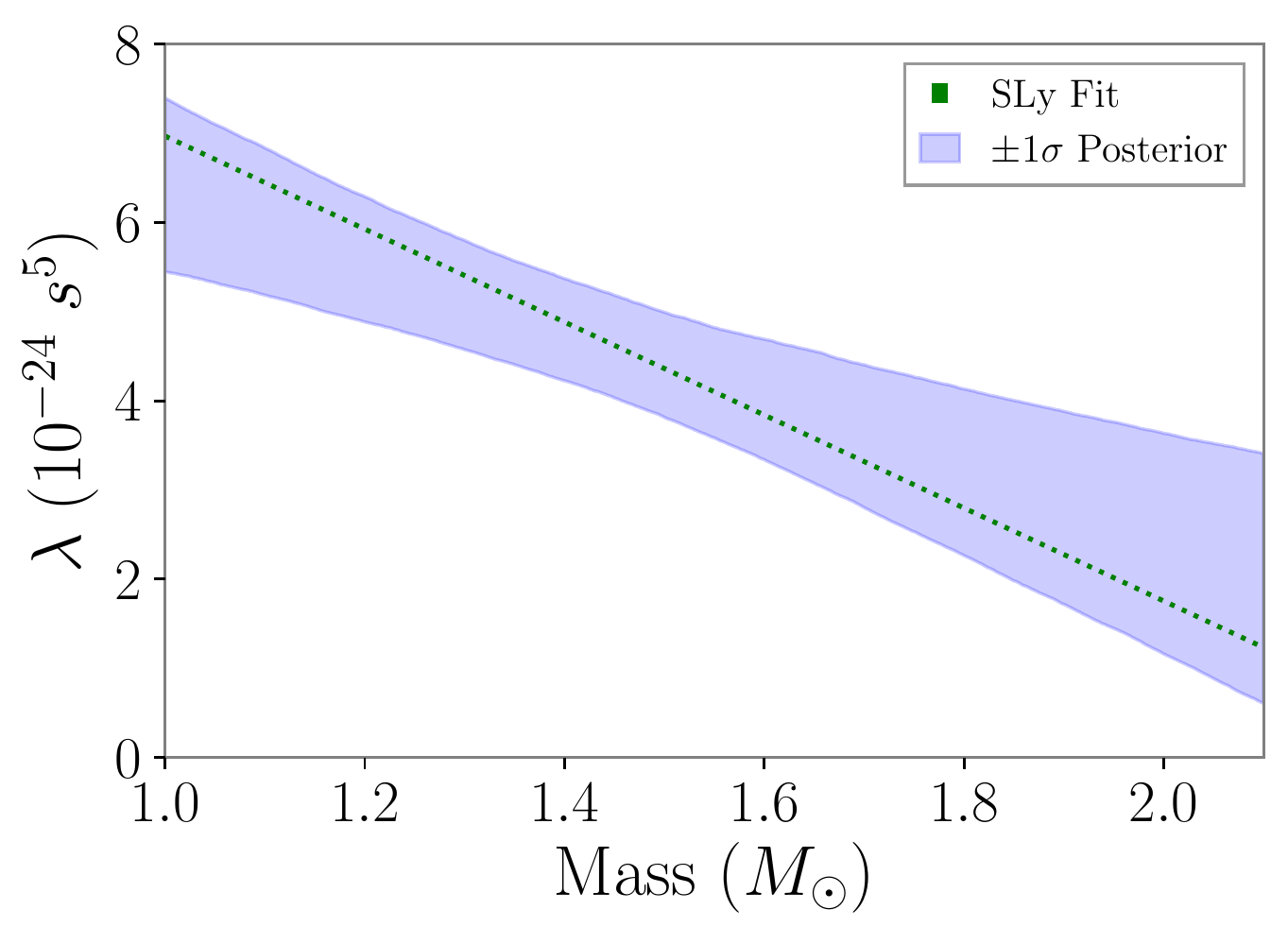}
    \includegraphics[width=\columnwidth]{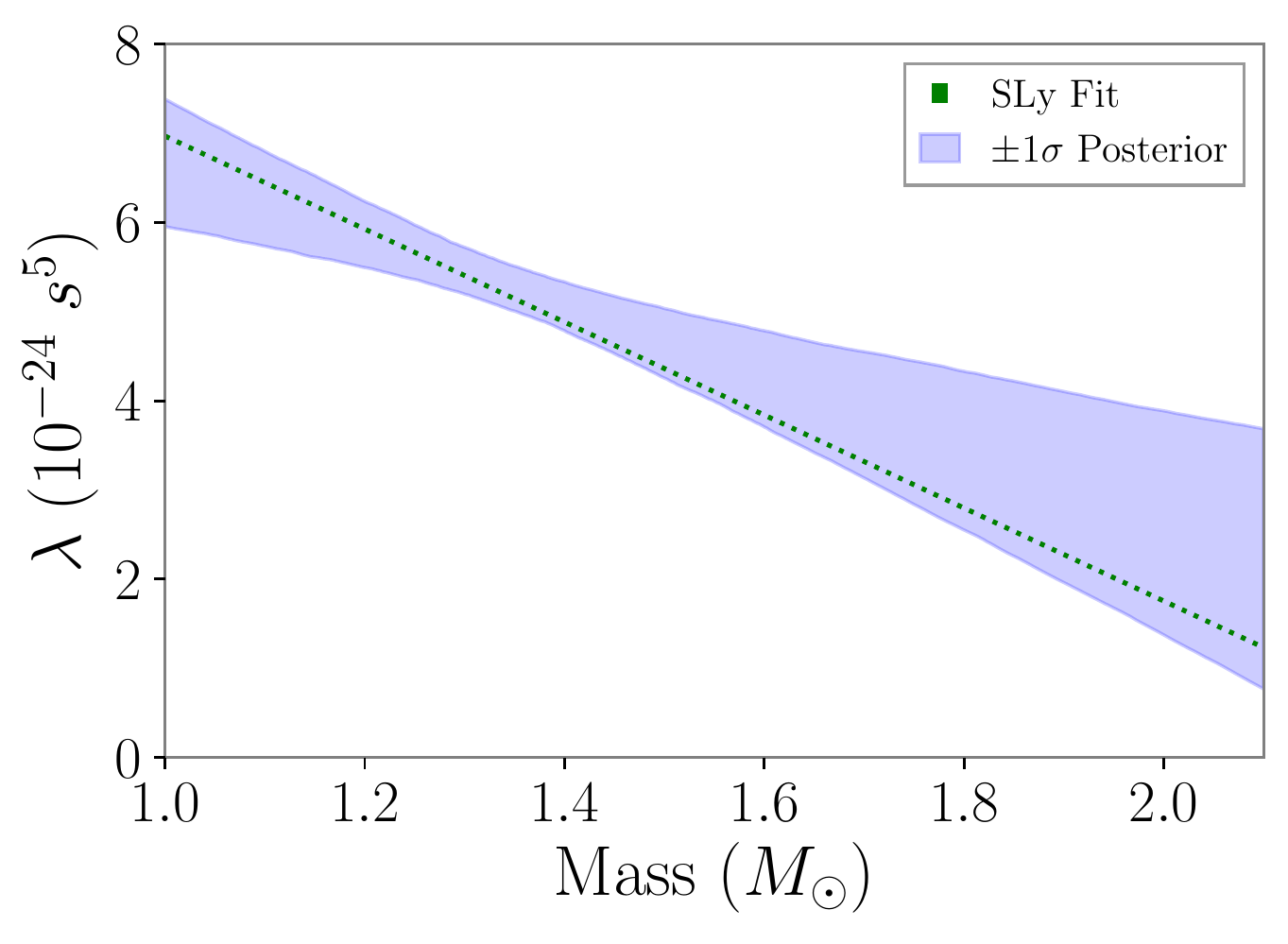}
    \caption{Inferred $\lambda - m$ parameter space from the analysis using only low SNR events (left) and high SNR events (right). Note the better recovery of EOS parameters from including few high SNR events compare to many low SNR events.}
    \label{fig:snrlambdas}
\end{figure*}

\begin{figure*}
    \centering
        \includegraphics[scale=0.6]{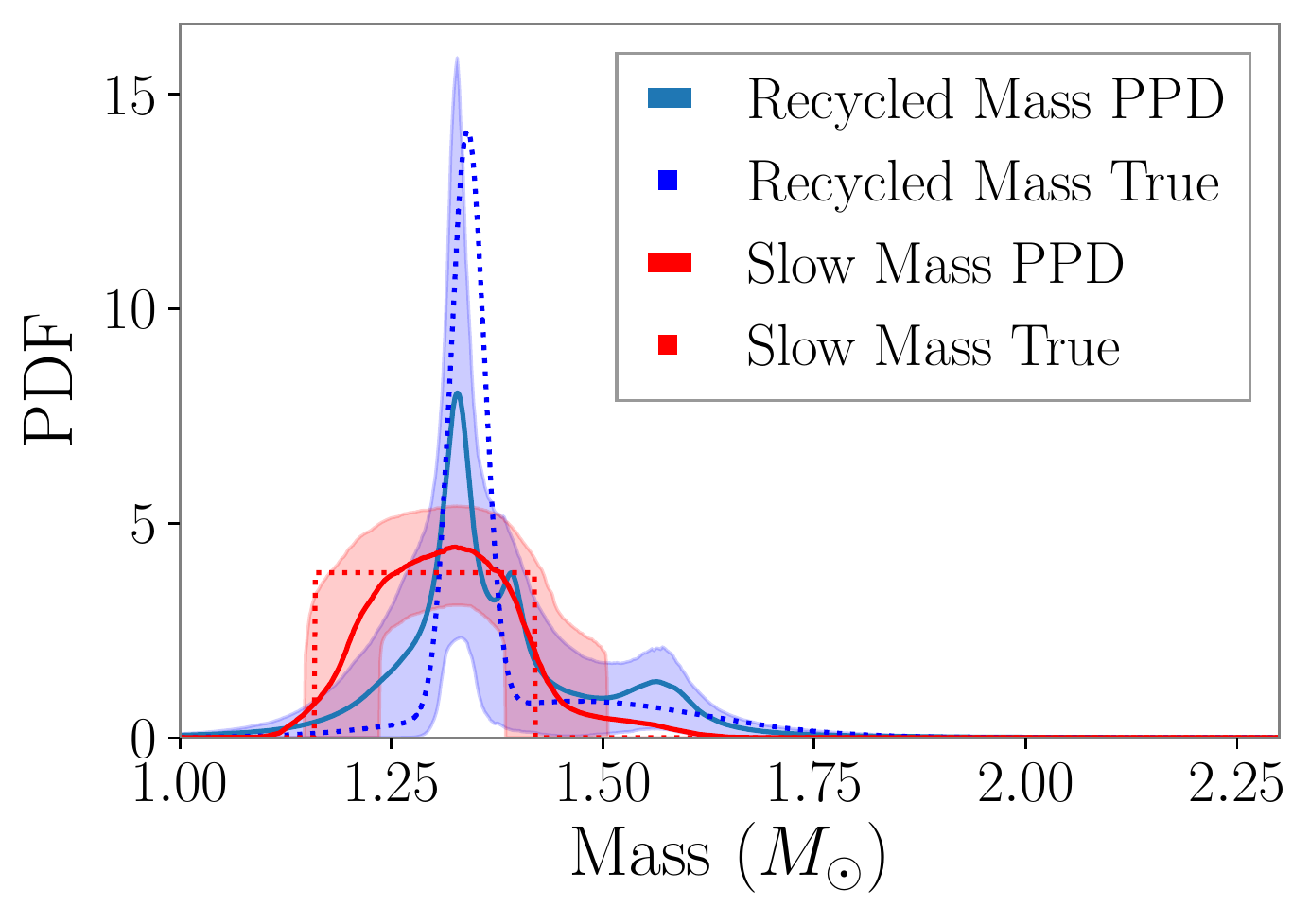}
    \centering
        \includegraphics[scale=0.6]{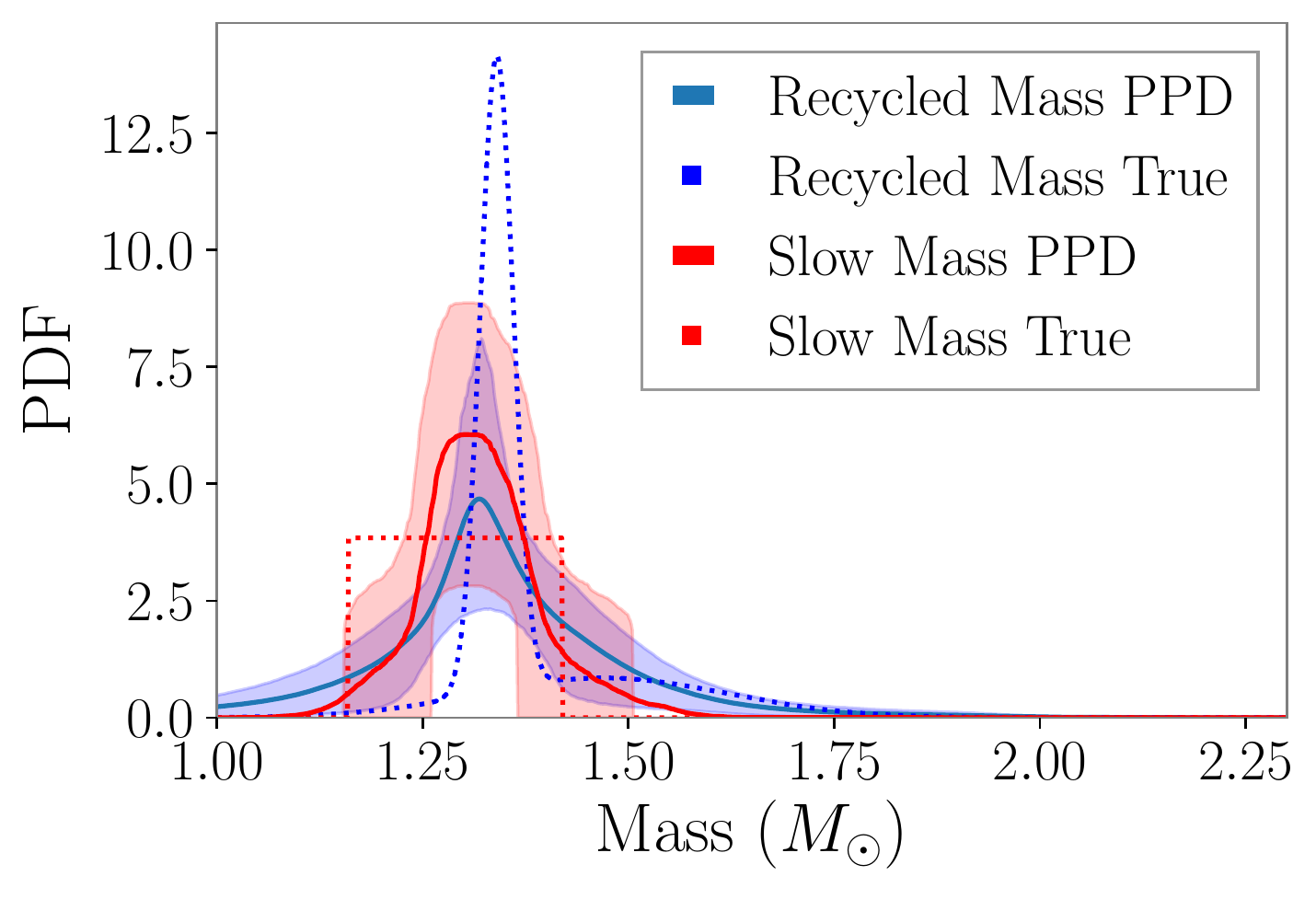}
    \caption{Inferred mass distribution from the full mass + EOS analysis of only low SNR events (left) and only high SNR events (right). Solid lines represent the posterior predictive distribution (PPD). The recycled (slow) distribution is colored blue (red), with shading representing the $\pm 1\sigma$ $(68\%)$ credible region from the posterior. Dashed lines show the input distribution. Compare to Figure \ref{fig:ppd} to note the worse recovery due to not including the full set of events.}
    \label{fig:snrppds}
\end{figure*}

\begin{figure}
    \centering
        \includegraphics[scale=0.6]{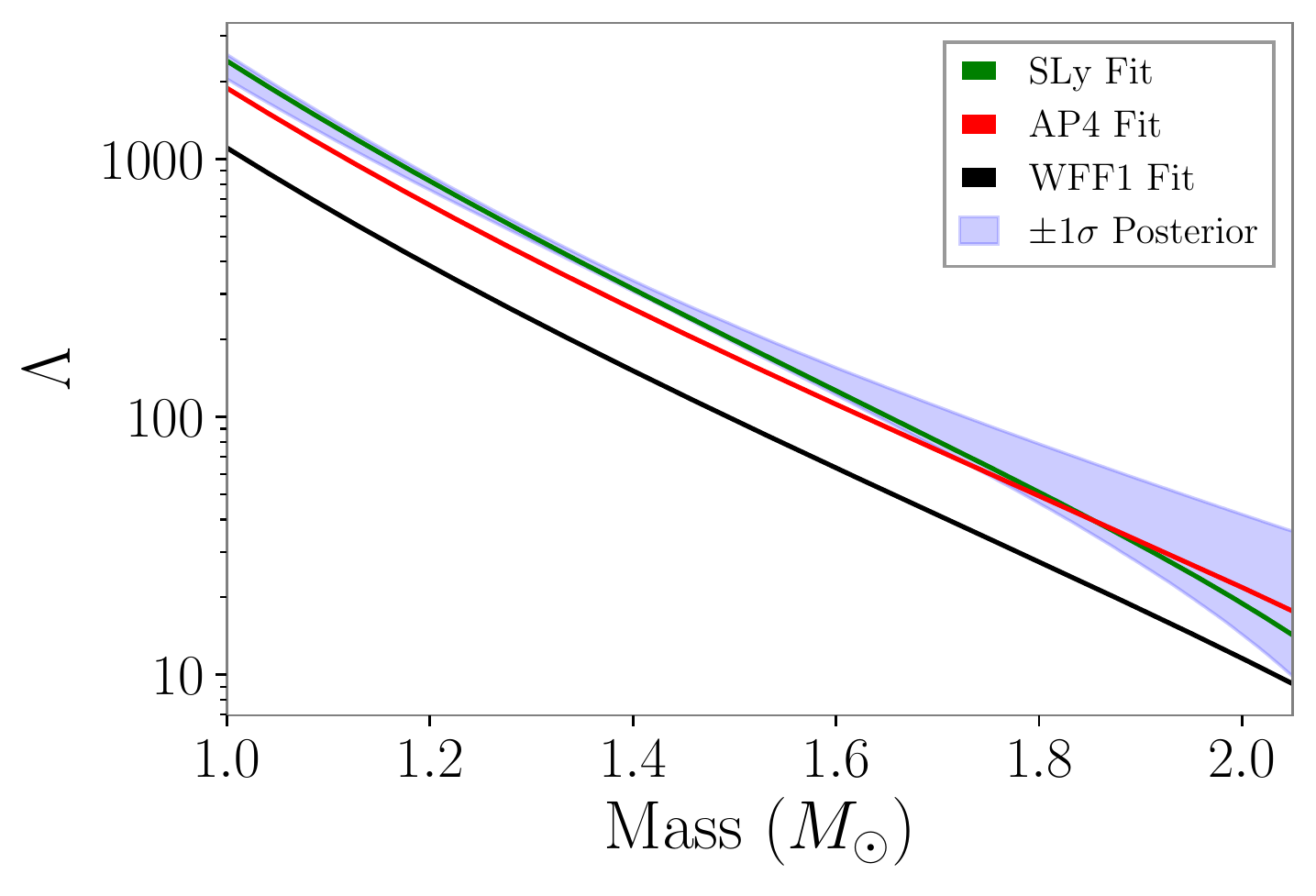}
        \includegraphics[scale=0.6]{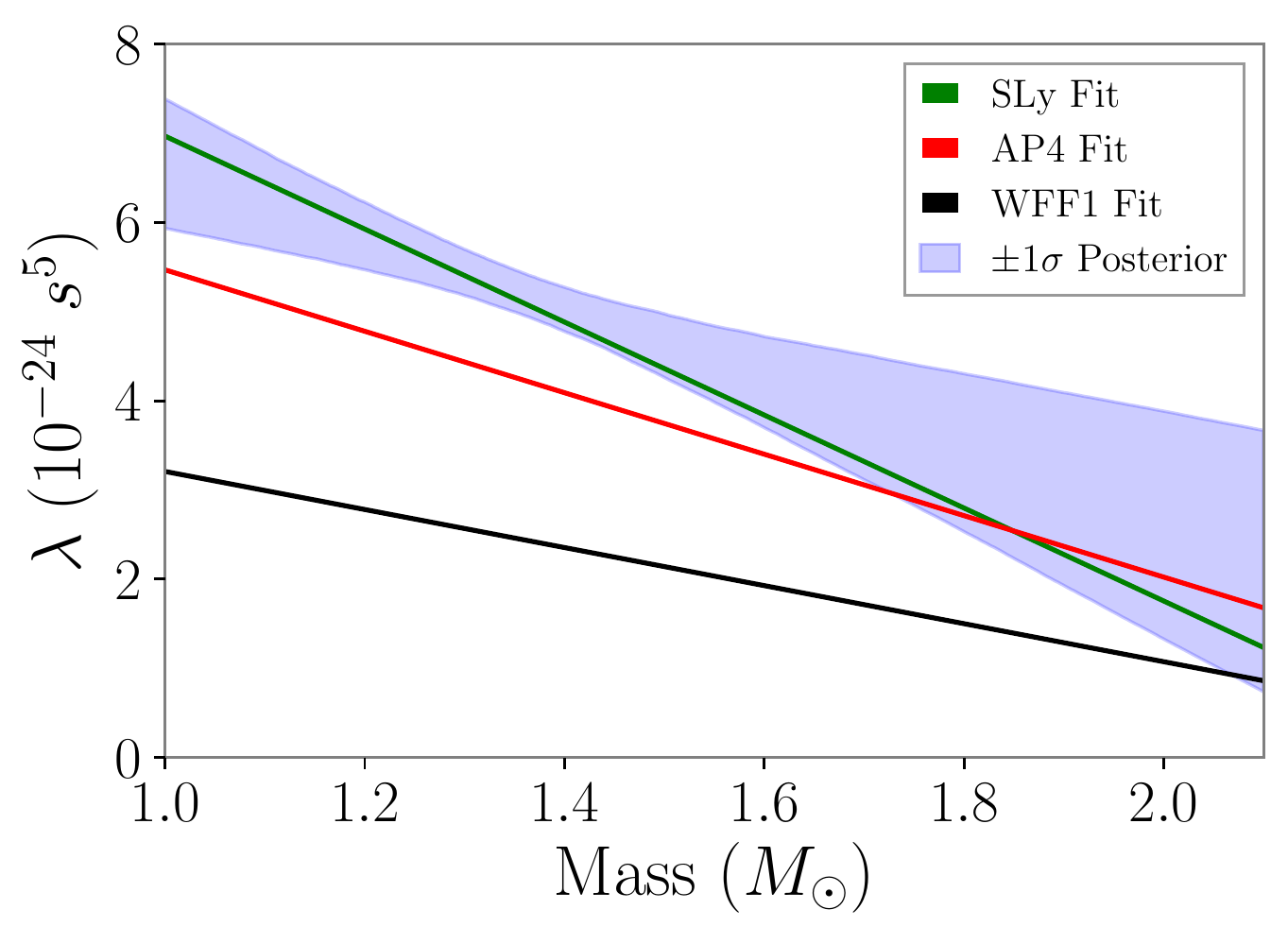}
    \caption{Inferred $\Lambda - m$ (top) and $\lambda - m$ (bottom) parameter space from the population + EOS analysis. Shaded region corresponds to $\pm 1\sigma$ $(68\%)$ region from $c_0$ and $c_1$ posterior samples. For reference, selected EOS curves overplotted.
    }
    \label{fig:compareLambda}
\end{figure}
Similarly, good recovery is also seen in the EOS parameters ($c_0$ and $c_1$), around the values input for our SLy fit which predicts $\Lambda(1.4 M_\odot)=313 $. We infer $\Lambda(1.4 M_\odot) = 322^{+27}_{-25}$, which is constrained to within $\sim$17\% of the true value from the SLy fit. We also recover $\Lambda(1 M_\odot) = 2282^{+435}_{-333}$ and $\Lambda(2 M_\odot) = 29^{+18}_{-23}$, the inferred dimensionless tidal deformability of a 1 $M_\odot$ and 2 $M_\odot$ neutron star, respectively. The relatively wide credible intervals for these parameters can be understood as a result of the vast majority of our simulated neutron stars having masses closer to $1.4 M_\odot$, with very little support in the mass distribution at $1 M_\odot$ or $2 M_\odot$.
For reference, we also plot the $\Lambda - m $ relationship for two other soft EOSs: AP4 \citep{ap4} and WFF1 \citep{wff1} which are EOSs compatible with observations from GW170817 \citep{gw170817}.
This way we show that the recovery can also favor the input EOS over some similar EOSs.
Thus, we will be able to robustly distinguish between these EOSs after observing 37 events.

As noted in Section \ref{Models}, the $M_{\max}$ parameter is fairly insignificant in this work, with its recovered posterior distribution relatively flat (see Figure \ref{fig:fullcorner}). There is however a sharp, slanted cutoff in the correlation plot between $c_1$ and $M_{\max}$, owing to the constraint we impose on $\Lambda(M_{\max})$ (see Section \ref{Models}).

The recovery of the EOS parameters obtained in this work stands in contrast to what was found in \cite{Agathos2015}. Simulating component masses from a narrow Gaussian peaked at $1.35 M_\odot$, but assuming a flat mass prior, the authors needed $\mathcal{O}(>100)$ events in their catalog to distinguish their candidate EOSs. In contrast to this work, those authors aimed to distinguish between soft, moderate, and stiff EOSs, whereas we only consider 3 similar soft EOSs, limiting the significance of a direct comparison between works. Despite comparing more similar EOSs in our work, we find a substantially lower threshold for distinguishing EOSs than in \cite{Agathos2015} by performing population inference simultaneously. 

Motivated by \cite{Lackey2015, Hernandez2019}, we test how well this method can constrain the mass distribution and EOS hyperparameters from only lower-SNR or higher-SNR events by repeating the above analysis, but limiting ourselves to a portion of the events.
The low SNR events are the 24 of our simulated events which had a network SNR $<$ 20 (but above our SNR threshold of 8), and the remaining 13 events are the events with SNR $>$ 20. 

Consistent with those studies, we find that EOS-sensitive parameters show definitively worse recovery when only including the low SNR events.
The inferred $\pm 1\sigma$ parameter space in the left panel of Figure \ref{fig:snrlambdas} in this case is much wider than when all the events are included, indicating that the SNR $>$ 20 events are providing a significant amount of the EOS information in this analysis. For instance, when only considering these low SNR events, we infer $\Lambda(1.4 M_\odot) = 309^{+61}_{-63}$. On the other hand, only analyzing the 13 events with SNR $>$ 20 provides $\Lambda(1.4 M_\odot) = 324^{+28}_{-29}$ (see right panel of Figure \ref{fig:snrlambdas}); constraints comparable to those from the full analysis with all SNR $>$ 8 events (see left panel of Figure \ref{fig:compareLambda}). 

As can be seen in Figure \ref{fig:snrppds}, the mass distribution recovery is poorer in the low-SNR case compared to including all events in the analysis. Specifically, we recover $\mu_{Ra} = 1.33^{+0.07}_{-0.11} M_\odot$; while this interval contains the true value of $\mu_{Ra}$, it is 3 times larger than what we get from the analysis using the full set of events. The 13 high SNR events contribute less information to the mass distribution inference than the 24 low SNR events, giving an inferred value $\mu_{Ra} = 1.30^{+0.07}_{-0.21} M_\odot$; a credible interval $\sim$4.5 times larger than that from the analysis using the full set of events. Therefore, unlike the case for the EOS, the mass distribution is not preferentially informed by high SNR events but is most sensitive to the number of events in the population. Because the mass parameters (chirp mass in particular) tend to be relatively well-constrained, several observations of lower-SNR BNS merger events can provide constraints on the mass distribution of merging neutron stars. 

To estimate the bias from not inferring mass distribution and EOS hyperparameters simultaneously, we conduct the analysis from above but only sample the mass distribution hyperparameters and make GMM density estimates of $\mathcal{M}$ and $q$ for our 37 observed events. By only considering the mass parameters in our analysis, we neglect the $\Lambda - m$ relationship from the EOS and implicitly (mis)model the tidal parameters as independent draws from the prior distribution used in the single-event sampling.
As seen in Figure \ref{fig:massonlyppd}, the mass distribution becomes noticeably biased at the 68\% credible level \footnote{The inferred distribution is consistent with the input distribution at the 90\% credible level, however.}, with the PPD of the mass spectrum shifted from the input distribution. In this case, we recover $\mu_{Ra} = 1.32^{+0.02}_{-0.09} M_\odot$, a credible interval which almost does not contain the true value. This bias is due to ignoring the correlations between the mass parameters (particularly the mass ratio) and the tidal parameters, which can be seen in the 2d posterior panels between $q$ and $\tilde{\Lambda}$ in single-event posteriors of our simulated events (see Figure \ref{fig:simulated_gmms}).
Therefore, inferring the $\Lambda - m$ relationship simultaneously with the mass distribution is necessary for an unbiased result.

\section{Discussion}\label{Discussion}

In this work, we demonstrate a new method of hierarchically combining posterior distributions from BNS merger events and inferring mass distribution and EOS parameters simultaneously.
The initial step of using GMM density estimates in our transformed space reliably reflects the observed posterior distribution and allows for the evaluation of single-event likelihoods at arbitrary points in parameter space for arbitrary subsets of single-parameters efficiently.

We show that our method can recover underlying population model parameters when combining BNS events simulated with realistic observational parameters and noise realizations, while also constraining parameters of the neutron star EOS.
Using our new method, we confirm the importance of inferring EOS and mass distribution parameters simultaneously to avoid potential bias in both the inferred mass distribution and EOS.

Additionally, we observe that both low-SNR ($<$ 20) and high-SNR ($>$ 20) observations contribute to mass population inference, with the few high-SNR observations providing the bulk of the EOS constraints.
This finding is generally consistent with the work in \citep{Lackey2015, Hernandez2019}.
Summarized in Figure \ref{fig:compareLambda}, the EOS recovery from the 37 simulated observations constrains the $\Lambda - m$ parameter space around the input EOS. 

The fourth observing run of the LIGO-Virgo-KAGRA network is expected to begin no earlier than Summer 2022 and last one year.
At the targeted upgraded sensitivity, it is estimated that there will be $10^{+52}_{-10}$ BNS detections, significantly raising the prospects for providing constraints on neutron star EOS and population models \citep{observing_prospects}.

The method presented in this work is generalizable to arbitrary population models and can incorporate parameterized models linking population observables to other single-event observables (i.e. $\Lambda - m$ relationship).
Gravitational wave population analysis using mass and spin models (see \cite{gwtc2}) could be similarly evaluated using this method by making GMM density estimates of mass and spin parameters, and sampling from the population models used in those studies.

We anticipate that the transformed Gaussian mixture model density estimation method employed here has additional potential applications, as it is robust to edge effects and has superior scaling with dimensionality compared to KDE and GP methods.
In addition to being able to handle the delta-function population model for tidal parameters, this method can be applied to any situation where the uncertainty in individual measurements is much larger than the domain of the population model, e.g., spin distributions that are not probeable with the method currently employed by the LIGO/Virgo collaboration analyses~\citep{gwtc1, gwtc2} (see also \cite{Wysocki2020}).
Further, Gaussian mixture models are a generative model and can therefore be used to generate $\mathcal{O}(10^{5})$ additional samples per second from the posterior distribution or as a proposal distribution for subsequent MCMC reanalyses building on the methods in~\cite{kombine, Ashton2021}.

While this proof-of-concept study used a simple toy model for the neutron star $\Lambda-m$ relation, more sophisticated models can be folded into the method.
Additionally, this method can be extended to include a model for the distribution of neutron star spins. Further, this study has focused on the situation when there are tens of measurements, the current population of binary neutron star systems is limited to two. In this small population regime the specific choice of population model/prior will significantly impact the inference.
We leave these extensions to future work.

\section{Acknowledgments}
We would like to thank Katerina Chatziioannou and Alan Weinstein for useful discussions. We would also like to thank Stefano Rinaldi for useful comments on the manuscript. Finally, we thank the anonymous reviewer for helpful suggestions and critiques on this manuscript. JG and CT acknowledge the support of the National Science Foundation, and the LIGO Laboratory. LIGO was constructed by the California Institute of Technology and Massachusetts Institute of Technology with funding from the National Science Foundation and operates under cooperative agreement PHY-1764464. This paper carries LIGO Document Number LIGO-P2100215.

The authors are grateful for computational resources provided by the LIGO Laboratory and supported by National Science Foundation Grants PHY-0757058 and PHY-0823459. This research has made use of data, software and/or web tools obtained from the Gravitational Wave Open Science Center (https://www.gw-openscience.org/) \citep{gwosc}, a service of LIGO Laboratory, the LIGO Scientific Collaboration and the Virgo Collaboration.

\bibliography{bibliography}
\bibliographystyle{aasjournal}

\appendix

\section{Single-Event Posteriors}
In this Appendix, we show the posterior distribution and samples drawn from the GMM fits for a range of our simulated events. By overplotting the samples from the posteriors and the GMM, we show that the GMM accurately characterizes the posteriors of individual events. We also note that the GMM is able to fit various features including peaks, correlations, and skews that appear in the transformed posterior distributions. This demonstrates the strength of using this density estimate as an analytic approximation of the likelihoods.

\begin{figure*}[t!]
    \centering
        \includegraphics[scale=0.35]{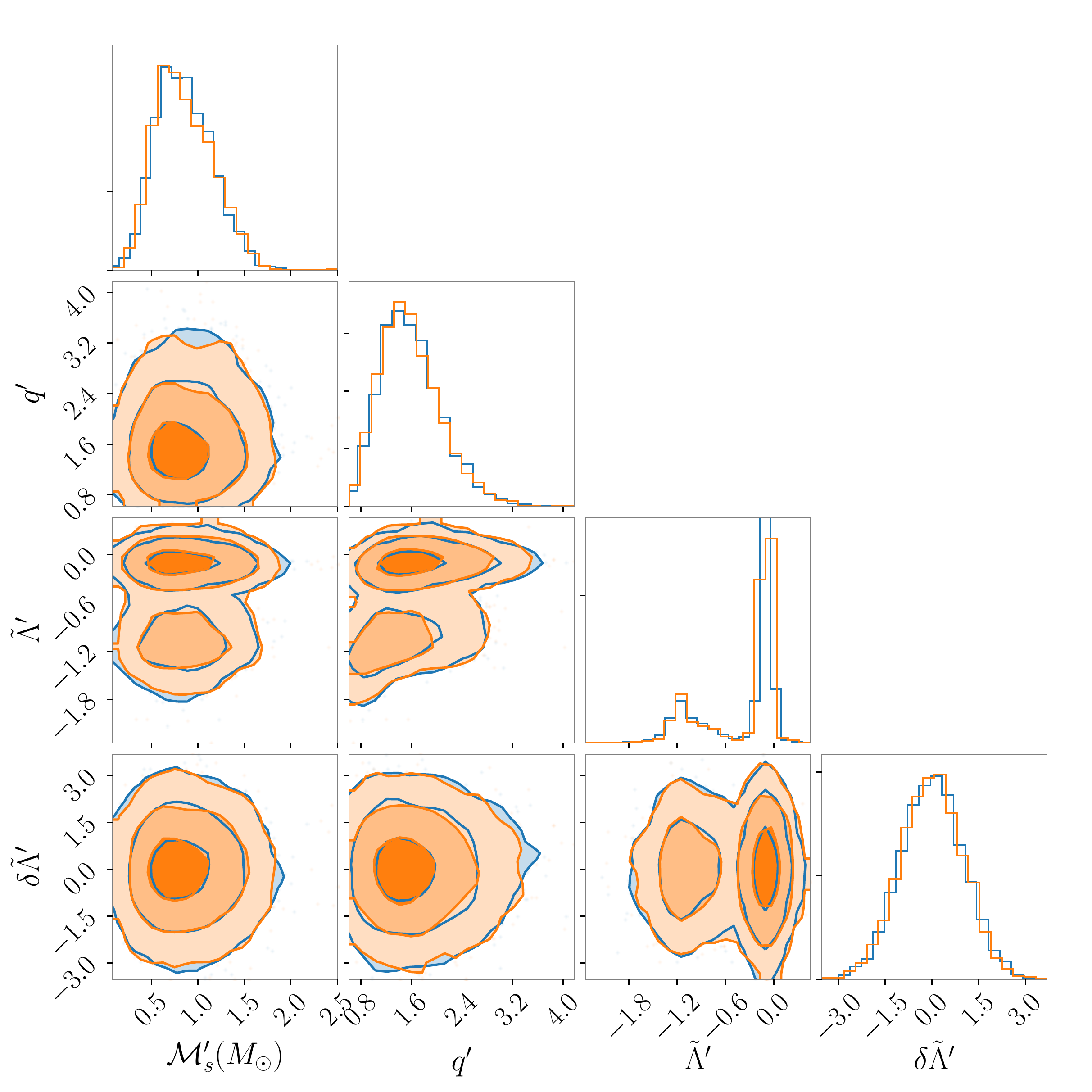}
    \centering
        \includegraphics[scale=0.35]{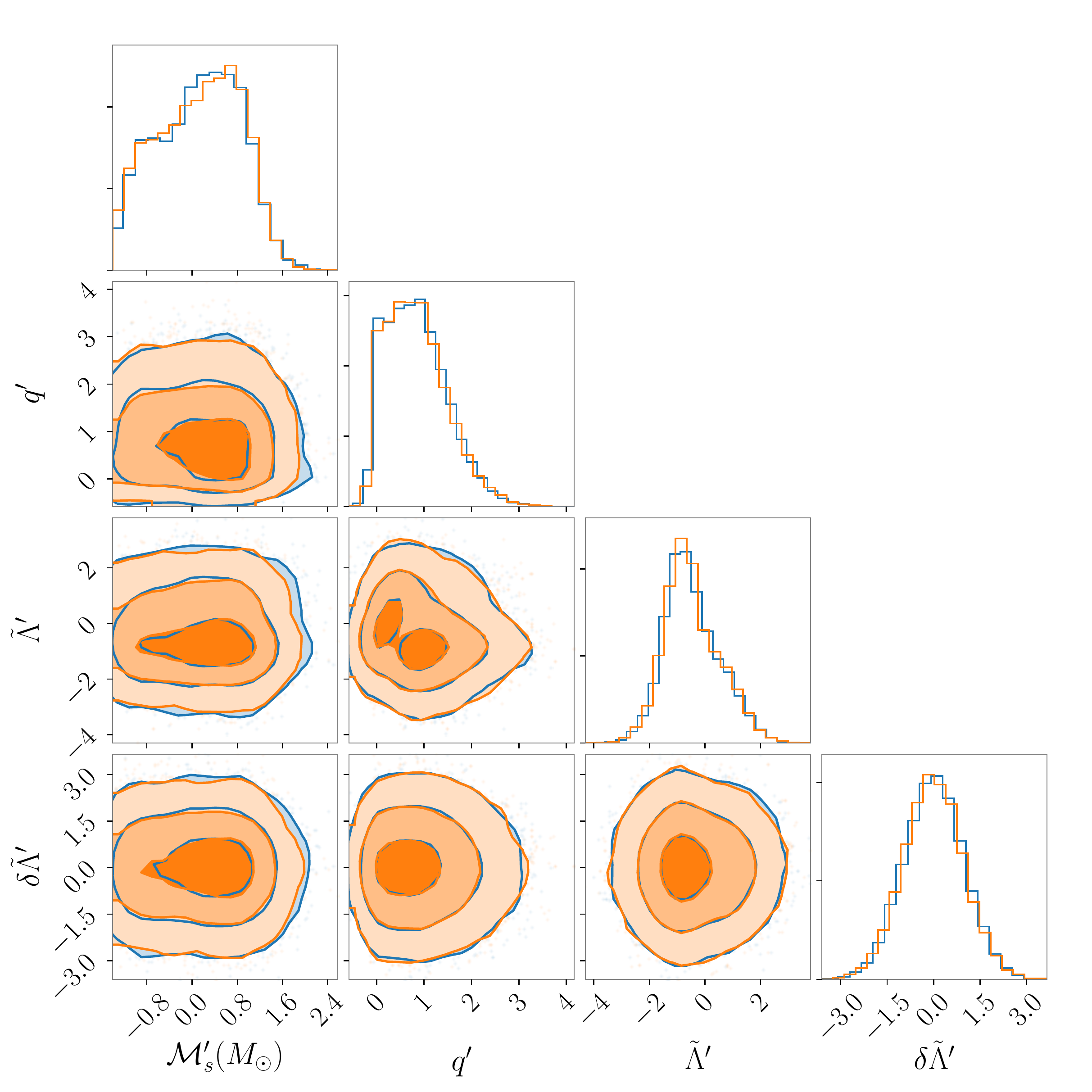}
    \centering
        \includegraphics[scale=0.35]{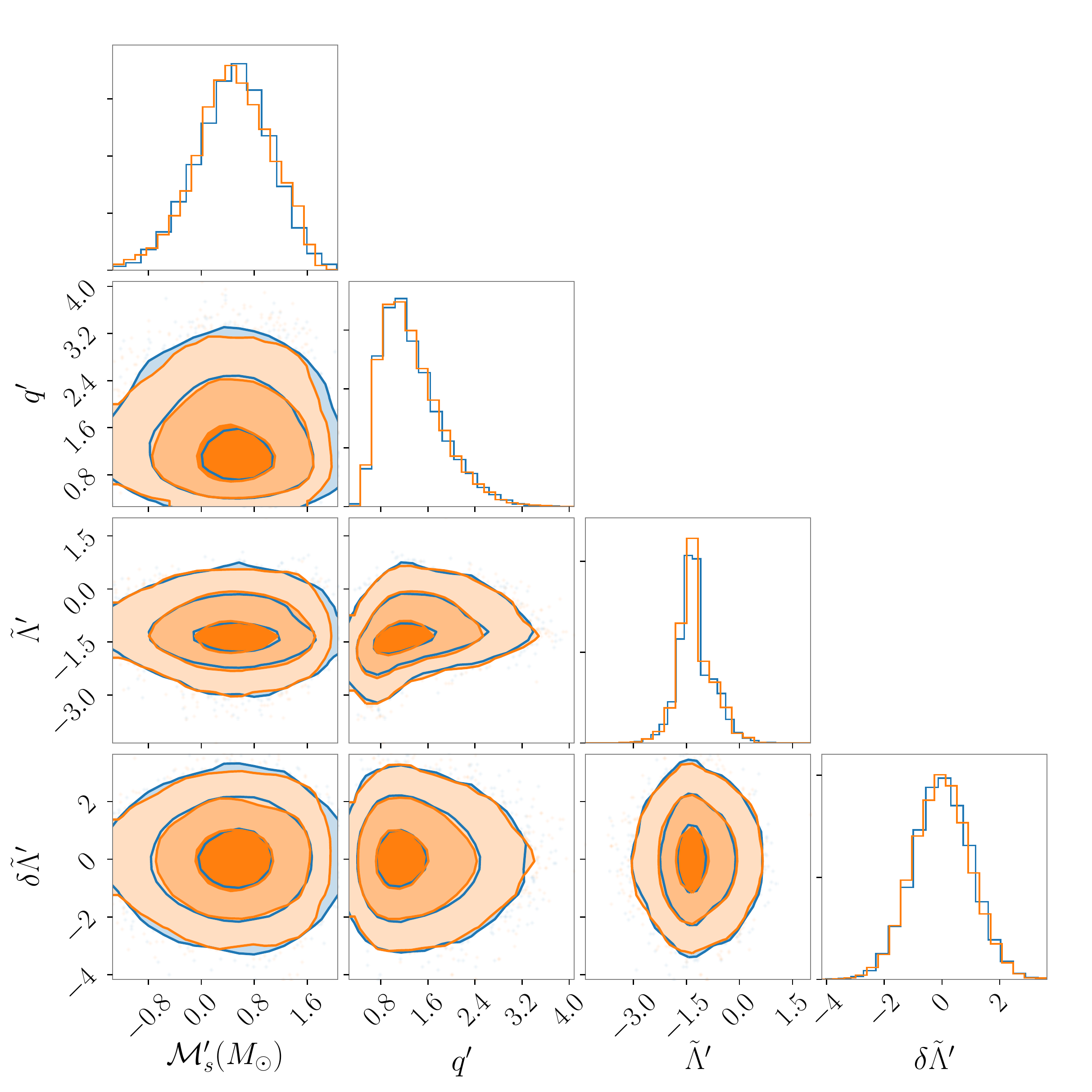}
    \hfill
    \centering
        \includegraphics[scale=0.35]{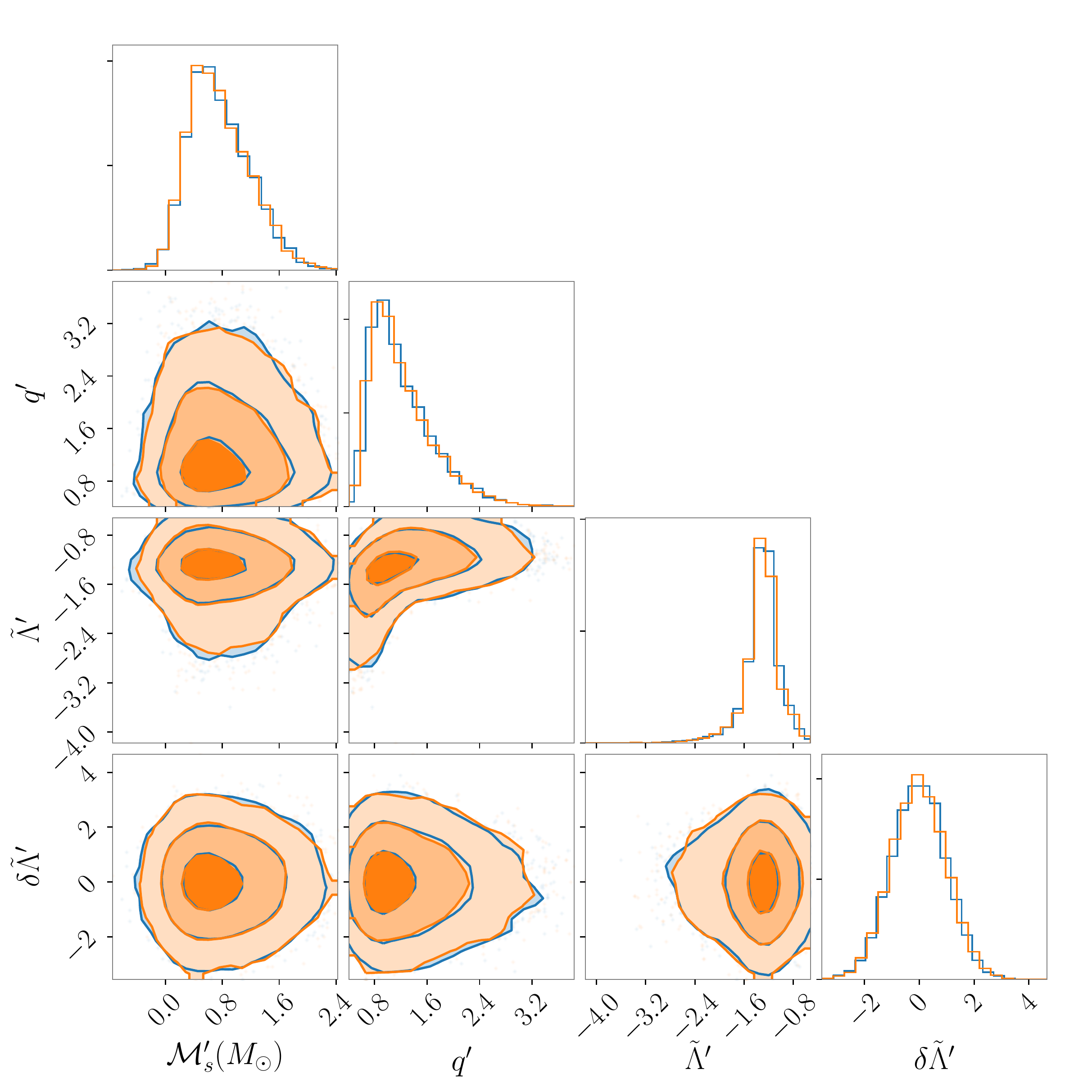}
    \caption{Posterior distributions of selected simulated events. Transformed samples are colored blue, and samples from the Gaussian mixture model density estimates are in orange. The overlap and consistency indicates that GMMs provide a good it in transformed space. Contours correspond to standard deviations in 2D space, such that 1-$\sigma$, 2-$\sigma$, 3-$\sigma$ contours are 39\%, 86\%, and 99\% confidence levels, respectively.}
    \label{fig:simulated_gmms}
\end{figure*}

The posterior distributions also show the correlations between the tidal parameters and mass parameters (center panels in Figure \ref{fig:simulated_gmms}). If the mass distribution is inferred independently from the tidal parameters, the information in the correlation (other than the $\mathcal{M} - q$ plot) is lost, impacting the recovery of the mass distribution.

\section{Hyper-Posterior}
\begin{figure*}
    \centering
        \includegraphics[scale=0.45]{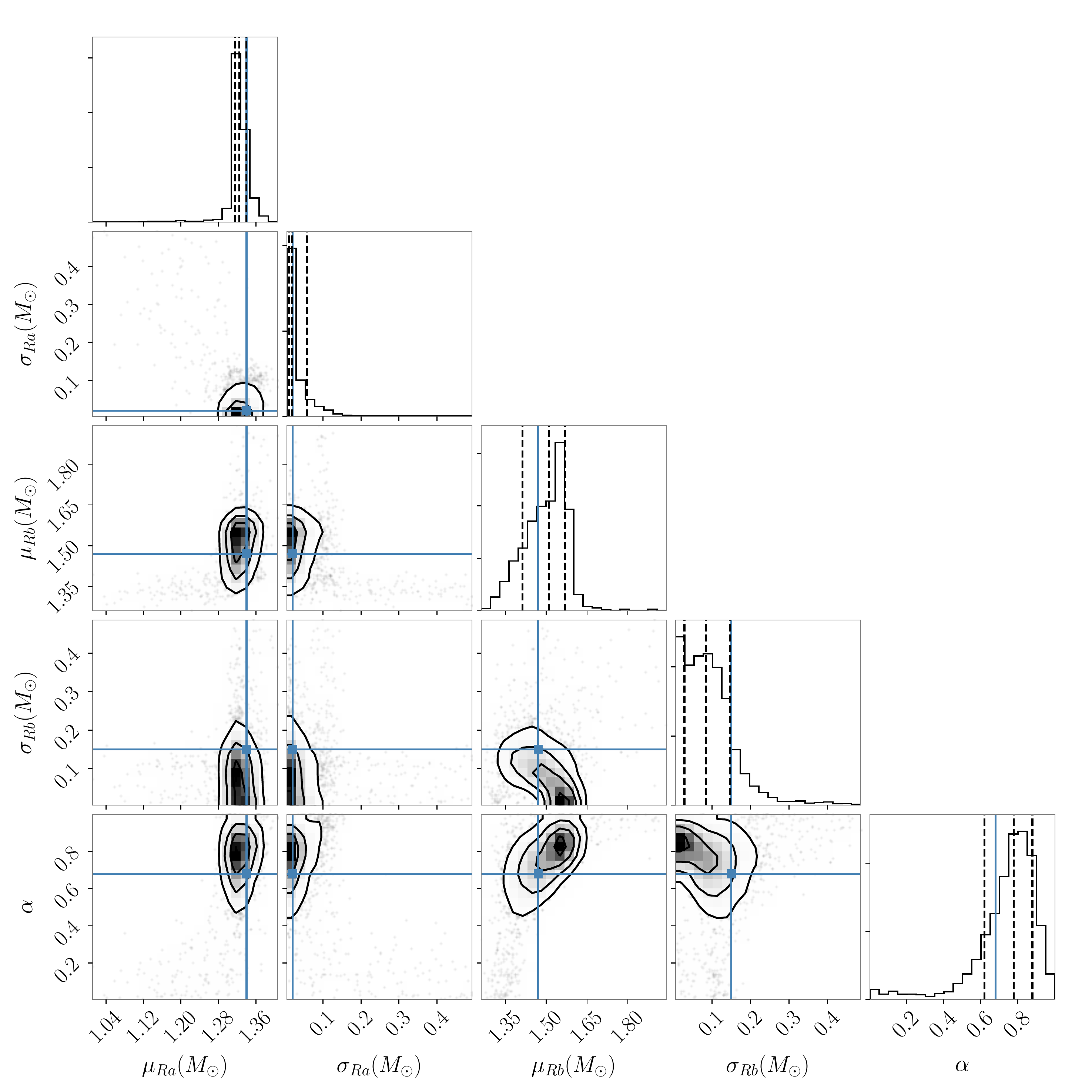}
    \centering
        \includegraphics[width=0.35\linewidth]{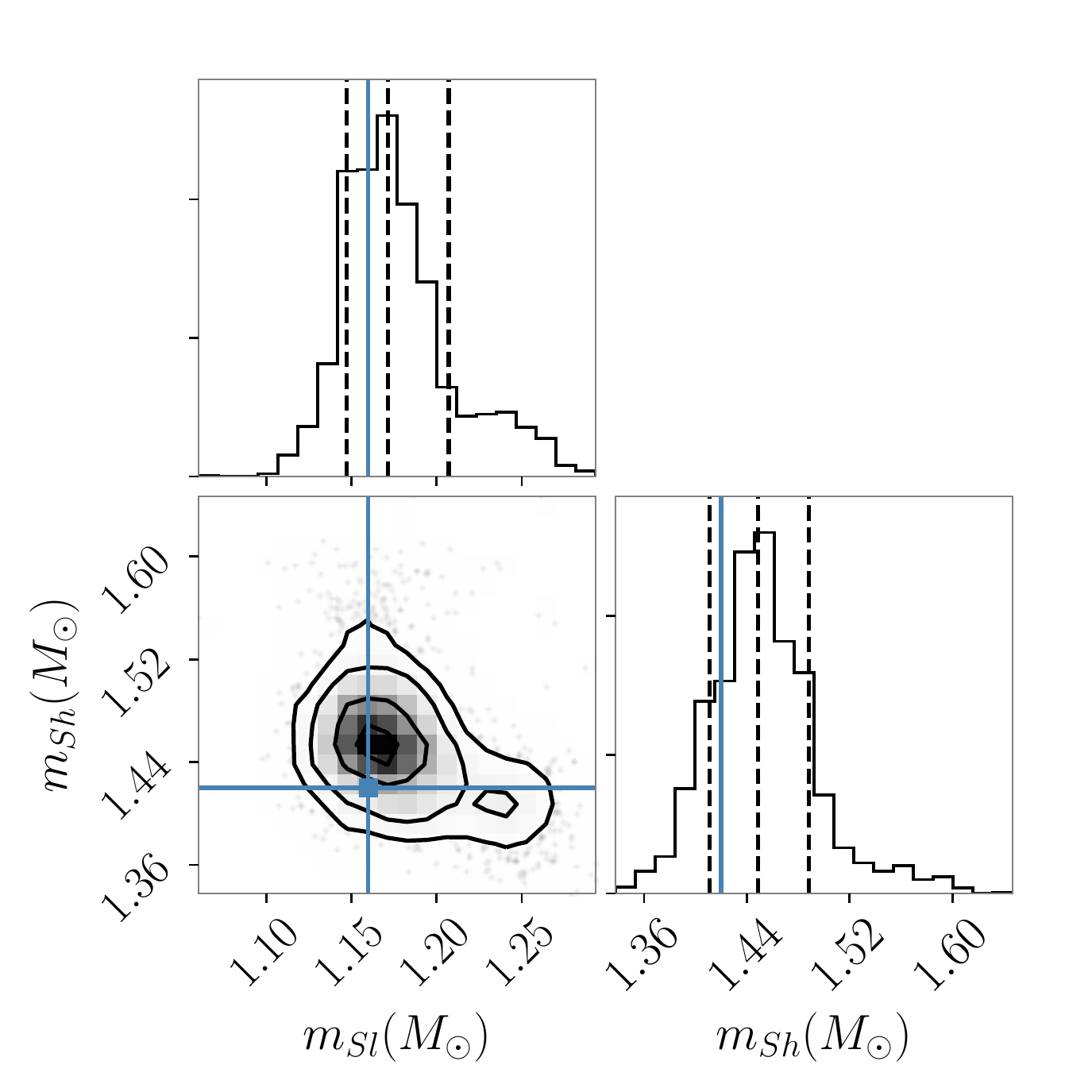}
        \includegraphics[width=0.5\linewidth]{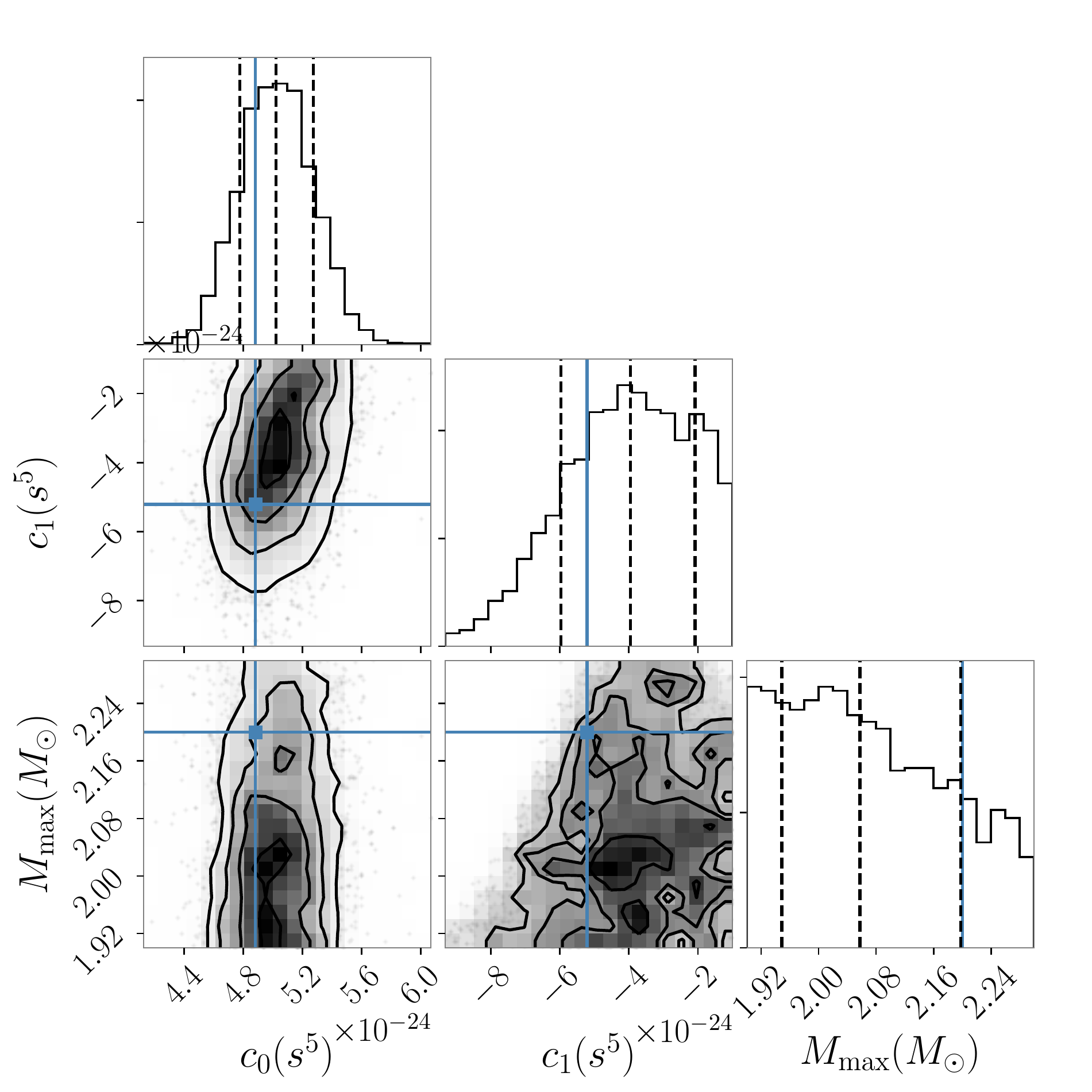}
    \caption{Inferred hyperparameter distributions for the recycled mass distribution hyperparameters (top), slow mass distribution hyperparameters (bottom left), and EOS hyperparameters (bottom right). Contours correspond to standard deviations in 2D space, such that 1-$\sigma$, 2-$\sigma$, 3-$\sigma$ contours are 39\%, 86\%, and 99\% confidence levels, respectively. See Table \ref{tab:params} for priors on each parameter.}
    \label{fig:fullcorner}
\end{figure*}

In Figure \ref{fig:fullcorner} we plot the recovered posterior distribution for the mass distribution and EOS hyperparameters inferred from the analysis using the 37 simulated events.

\section{Convergence of Monte Carlo Integrals}

Population analyses such as those presented here rely on the use of Monte Carlo integrals to marginalize over the single-event parameters, either by summing over posterior samples for each event with some fiducial prior, e.g.,~\cite{gwtc2}, or by summing over samples from the population model as in this work.
While such Monte Carlo integrals are asymptotically unbiased estimators, for a finite number of samples there is a finite uncertainty.
This uncertainty is generally neglected in the literature, although it has been discussed for the integral estimating the sensitivity function,  Equation~\ref{pdet}~\citep{Farr2019}.
For a generic Monte Carlo integral of some function $f$ over some set of $K$ samples $x_k$ the expectation is
\begin{equation}
    \left< f \right> = \frac{1}{K} \sum^{K}_{k} f(x_k)
\end{equation}
and the fractional uncertainty is
\begin{equation}\label{eq:per-event-variance}
    \left(\frac{\Delta f}{\left< f \right>}\right)^2
    = \frac{1}{K} \frac{ \left< f^2 \right> - \left< f \right>^2 }{ \left< f \right>^2 }
    = \frac{\sum^{K}_{k} f^2(x_k)}{\left( \sum^{K}_{k} f(x_k) \right)^2} - \frac{1}{K}.
\end{equation}
We include the two notational forms to highlight the asymptotic form (center) and the practical method for evaluating the quantity (right).
From the asymptotic form, we note that if the moments of $f$ can be analytically computed, we would have an expected variance that is exactly inversely proportional to the number of samples being averaged over.

Neglecting the uncertainty in the estimate of the selection function, the uncertainty in our total likelihood is the logarithm of the product of many Monte Carlo integrals, the standard rules of propagating uncertainties yield
\begin{equation}\label{deltalnL}
    \Delta \ln \mathcal{L}(\{d_{i}\} | \Omega) = \sqrt{\sum^{N}_{i} \left(\frac{\Delta \mathcal{L}_{i}}{\left< \mathcal{L}_{i} \right>}\right)^2}.
\end{equation}
Here, the quantity in the sum is the equivalent of Equation~\ref{eq:per-event-variance}.
We emphasize that we are interested in the absolute uncertainty in the log-likelihood.This uncertainty will increase with the number of events for fixed per event variance.
In order to maintain a constant uncertainty, the required number of samples per hyperparameter is proportional to the number of events.
Thus the total number of samples required for constant uncertainty scales like $N^2$.

\subsection{Population sample weighting}
In this work, we draw samples from the population model and evaluate Equation~\ref{likelihoodgmm}, i.e., $f(x_k) \rightarrow \mathcal{L}_i(\theta')$.
In addition to estimating the statistical uncertainty, we note that we can directly sample the distribution of $\ln\mathcal{L}$ by repeatedly evaluating the likelihood with different realizations of samples from the population model.

\begin{figure*}
    \centering
    \includegraphics[scale=0.8]{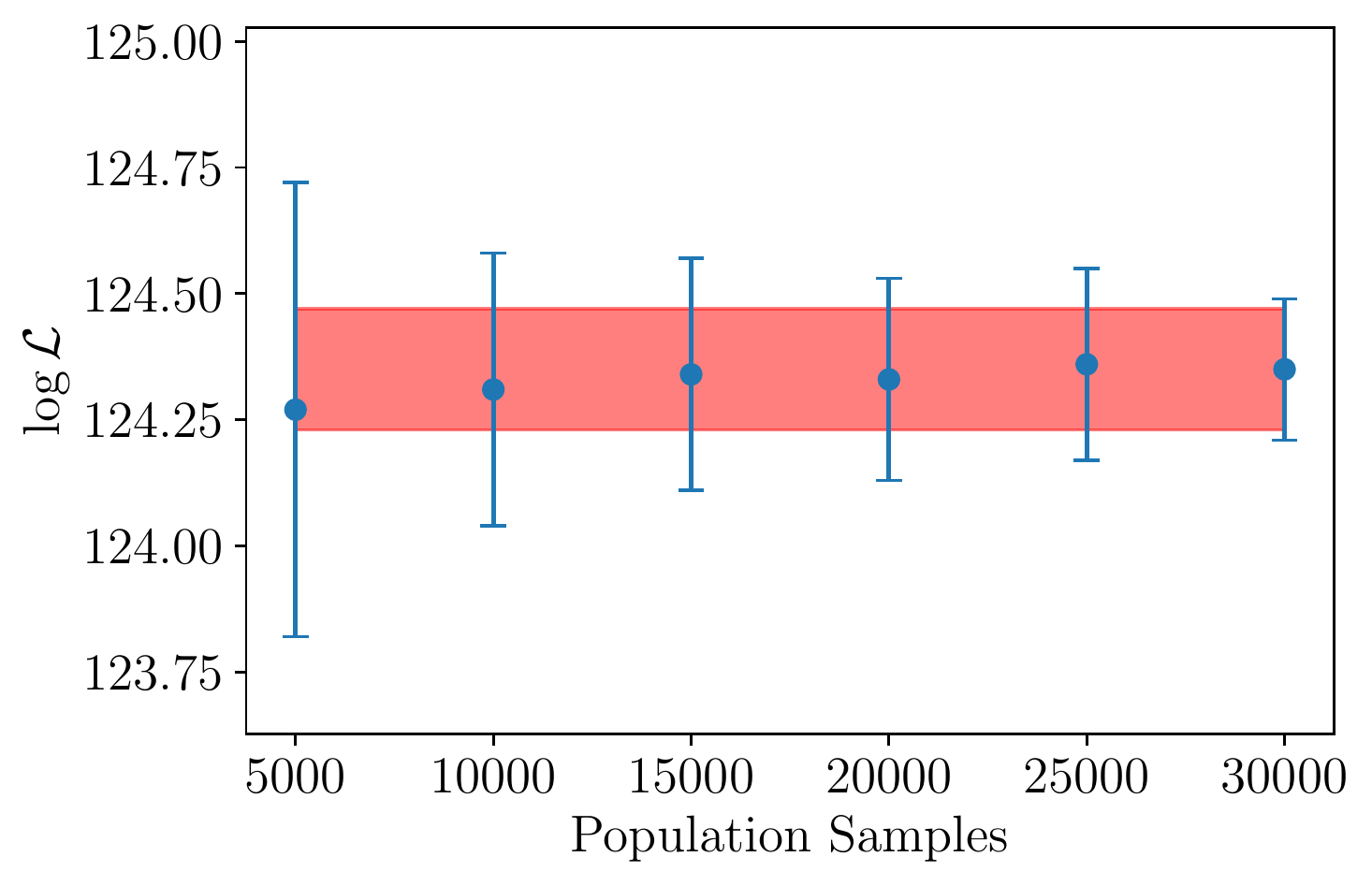}
    \caption{Log likelihood (Equation \ref{likelihood}) for the true population hyperparameter values as a function of the number of samples $M$ from the population. Each data point is an average over 100 likelihood iterations and error bars are $\pm 1\sigma$ $(68\%)$ uncertainties. By $M=15,000$, the Monte Carlo integration is stable and increased values of $M$ only increase the precision marginally. The red shaded region corresponds to the $68\%$ net uncertainty resulting from using Monte Carlo integration via reweighting the single-event posterior samples in the mass population model (see Equation \ref{eq:per-event-variance})}.
    \label{fig:converge}
\end{figure*}

In Figure \ref{fig:converge} we plot the average log likelihood and $1\sigma$ uncertainty for 100 trials as a function of increasing number of population samples $M$ (blue). This provides a test of convergence of the Monte Carlo integration, as a converged integral should be invariant under changes to the number of samples $M$. We note that using $M=15,000$ samples, with an associated $\Delta\ln\mathcal{L} = 0.23$, is sufficient for convergence of this likelihood integration. In order to confirm that this integral is well-behaved for our choice of $M$, we perform a check by computing the likelihood again for each hyperparameter sample in our posterior distribution and reweight our original posterior distribution by this new distribution. We do this ten times to get ten new mock realizations of our posterior distribution. Differences between each simulated realization should therefore be due to statistical uncertainty when computing the Monte Carlo integral over random samples from the population distribution. We confirm qualitatively that these additional realizations are nearly identical to the original posterior distribution, indicating the Monte Carlo integral for the likelihood is stable for this choice of $M$.

\cite{Wysocki2020} further reduces the uncertainty in their implementation of the population sample reweighting method by sampling only from regions in the population model which have non-vanishing support for mass parameters in the single-event likelihoods. This increases the number of effective samples per Monte Carlo computation, presumably resulting in a reduced $\Delta\ln\mathcal{L}$ computed via Equation \ref{deltalnL}.

\subsection{Posterior sampling reweighting}

As a comparison, we compute the uncertainty in the calculated likelihood when reweighting single-event posterior samples in the likelihood.
In this case the quantity inside the sum is the ratio of the population model to the fiducial prior distribution $f(x_k) \rightarrow \pi(\theta | \Lambda) / \pi(\theta | \O).$
For this method, we are restricted to a single set of samples from the fiducial posterior distribution and so we must rely on the statistical uncertainty. We are also limited to only using posterior samples from the masses (no samples from the tidal parameters), as we cannot reweight posterior samples in a population model which accounts for a $\Lambda(m)$ relationship.
In Figure~\ref{fig:converge}, we show the uncertainty in an equivalent calculation in the red shaded region.
We center the region at the mean estimator using the population model sampling method with 30000 population samples for ease of comparison.
To calculate this uncertainty we use 4480 samples per event.
We reiterate here that, while this reweighting method may give less uncertainty in $\ln\mathcal{L}$ in this application (i.e., the single-event posteriors are much narrower than the population model), it cannot account for tidal effects in the population model, as the distribution for $\Lambda$ is a delta function for a given mass (i.e., the single-event posteriors are much wider than the population model). 
\end{document}